\documentclass[twocolumn, 10pt]{IEEEtran}

\usepackage{hyperref}
\usepackage{graphicx}
\usepackage{amssymb}
\usepackage{amsmath}
\usepackage{mathtools}
\usepackage{cite}
\usepackage{stfloats}
\usepackage{subfigure}
\usepackage{psfrag}
\usepackage[mathscr]{euscript}
\usepackage{acronym}  
\usepackage{booktabs}
\usepackage[table]{xcolor}
\usepackage{algorithm}
\usepackage[noend]{algpseudocode}

\acrodef{BS}{base station}
\acrodef{AS}{all service-coded caching}
\acrodef{OS}{one service-coded caching}
\acrodef{EC}{edge computing}
\acrodef{UE}{user equipment}
\acrodef{LoS}{Line-of-Sight}
\acrodef{NLoS}{Non-Line-of-Sight}
\acrodef{UDN}{Ultra-Dense Network}
\acrodef{SCA}{successive convex approximation}
\acrodef{PPP}{Poisson point process}
\acrodef{PP}{Poisson process}
\acrodefplural{PPP}[PPPs]{Poisson point processes}
\acrodef{MEC}{Mobile edge computing}
\acrodef{RAN}{radio access network}
\acrodef{KKT}{Karush-Kuhn-Tucker}
\acrodef{PDF}{probability density function}
\acrodef{CDF}{cumulative distribution function}
\acrodef{CCDF}{complementary cumulative distribution function}
\acrodef{SIR}{signal-to-interference ratio}
\acrodef{SINR}{signal-to-interference-plus-noise ratio}
\acrodef{PGFL}{probability generating functional}
\acrodef{ASE}{area spectral efficiency}
\acrodef{PGFL}{probability generating functional}
\acrodef{V2X}{vehicle-to-everything}
\acrodef{RGM}{random Gauss-Markov}
\acrodef{STP}{successful transmission probability}
\acrodef{RSU}{road side unit}
\acrodef{VUE}{vehicular user equipment}
\acrodef{IoT}{Internet of thing}
\acrodef{SC}{Service caching}
\acrodef{P-K}{Pollaczek-Khinchin}
\acrodef{JCS}{joint service caching decision and server splitting}
\acrodef{SSP}{successful service probability}
\acrodef{SUTP}{successful uplink transmission probability}
\acrodef{SDTP}{successful downlink transmission probability}
\acrodef{SCPP}{successful computation probability}
\acrodef{UCP}{uplink coverage probability}
\acrodef{DCP}{downlink coverage probability}
\acrodef{UC}{uniform service caching distribution}
\acrodef{PC}{proportional service caching distribution}
\acrodef{US}{uniform computation resource allocation}
\acrodef{PS}{proportional computation resource allocation}
\acrodef{VM}{virtual machine}
\acrodef{FIFO}{first input first output}
\acrodef{TDD}{time-division duplexing}
\acrodef{HTT}{Hyper-threading Technology}
\acrodef{w.r.t.}{with respect to}
\acrodef{SUM}{Successive upper-bound minimization}
\acrodef{DC}{difference of convex function}
\acrodef{CCCP}{concave-convex procedure}
\acrodef{LP}{linear programming}
\acrodef{CDF}{cumulative distribution function}
\acrodef{GP}{Geometric Programming}
\acrodef{UCPS}{\emph{Uniform service caching and proportional computation resource allocaiton}}
\acrodef{GCPS}{\emph{Geography-based service caching and proportional computation resource allocaiton}}
\acrodef{PCOS}{\emph{Popularity-based service caching and optimal computation resource allocation}}
\acrodef{TCPS}{\emph{Transmission performance-based service caching and proportional computation resource allocation}}
\acrodef{RT}{random service time}
\acrodef{DT}{deterministic service time}
\usepackage{color}
\usepackage{dsfont}
\usepackage{bbm}

\newtheorem{lemma}{Lemma}
\newtheorem{corollary}{Corollary}
\newtheorem{proposition}{Proposition}

\newtheorem{assumption}{Assumption}
\newtheorem{problem}{Problem}

\newcommand{\Adensity}[1]{\lambda_{\text{#1}}}
\newcommand{\Appp}[1]{\Phi_{\text{#1}}}

\newcommand{\MetricSSP}[2]{\mathcal{P}_k\left(\mathbf{#1},\mathbf{#2} \right)}
\newcommand{\MetricSSPC}[2]{\tilde{\mathcal{P}}_{k}\left(\mathbf{#1},\mathbf{#2} \right)}
\newcommand{\MetricSSPCC}[3]{\tilde{\mathcal{P}}_{\text{#1}}\left(\mathbf{#2},\mathbf{#3} \right)}
\newcommand{\Ametric}[1]{P_{n}^{(\text{#1})}}
\newcommand{\AmetricP}[1]{P_{n,j,#1}^{(\text{Q})}}

\newcommand{\ARate}[1]{R^{(\text{#1})}}
\newcommand{\ATime}[1]{D_{n}^{(\text{#1})}}
\newcommand{\ATimeQ}[2]{D_{n,#1}^{(\text{#2})}}
\newcommand{\Ivar}[1]{\mathbf{#1}^{(r)}}

\newcommand{\BIvar}[1]{\mathbf{#1}_k^{(r)}}
\newcommand{\Apower}[1]{P_{\text{#1}}}
\newcommand{\OptVar}[1]{\mathbf{#1}}
\newcommand{\Tsize}[1]{S_{n}^{(\text{#1})}}
\newcommand{\Tdead}[1]{\gamma_{n}^{(\text{#1})}}
\newcommand{\TCoverage}[1]{\beta_{n}^{(\text{#1})}}

\newcommand{\AMetricSSP}[1]{\mathcal{P}_{\infty}\left(#1 \right)}

\newcommand{\appsubAgen}[1]{\tilde{\mathcal{F}}_{\mathbf{a},#1}\left(\mathbf{a},\mathbf{a}_{#1}^{(r)},\mathbf{b}_{#1}^{(r)}\right)}
\newcommand{\subBgen}[1]{\mathcal{F}_{\mathbf{b}_j,#1}\left(\mathbf{b}_j,\mathbf{a}_{#1}^{(r)}\right)}
\newcommand{\appsubBgen}[1]{\tilde{\mathcal{F}}_{\mathbf{b}_j,#1}\left(\mathbf{b}_j,\mathbf{a}_{#1}^{(r)},\mathbf{b}_{#1}^{(r)}\right)}
\newcommand{\IOptsol}[2]{\bar{#1}_{#2}^{(r+1)}}
\newcommand{\appsubT}[1]{\tilde{\mathcal{P}}_{#1}\left(\mathbf{T},\mathbf{T}^{(r)} \right)}
\newcommand{\OptsubAspeR}[1]{\bar{\mathcal{H}}_{j,\text{#1}}\left(\mathbf{T}^*\right)}
\newcommand{\subAspeR}[1]{\mathcal{H}_{j,\text{#1}}\left(\mathbf{b}_j,\mathbf{T}^*\right)}

\newcounter{eqnback}

\begin{document}
\bstctlcite{IEEEexample:BSTcontrol}
\newcommand{\red}[1]{{\textcolor[rgb]{1,0,0}{#1}}}
\newcommand{\blue}[1]{{\textcolor[rgb]{0,0,1}{#1}}}
\newcommand{\green}[1]{{\textcolor[rgb]{0.184,0.616,0.153}{#1}}}

\newcommand{\paperTitle}{Joint Service Caching and Computing Resource Allocation for Edge Computing-Enabled Networks}
%
%
 
 

\title{\paperTitle}

\author{
	\vspace{0.2cm}
	Mingun~Kim, 
	Hewon~Cho,
	Ying~Cui, \emph{Member}, \emph{IEEE}, and 
	Jemin~Lee, \emph{Member}, \emph{IEEE}
	%
	%
	%
	\thanks{
		The material in this paper was presented, in part, at the IEEE Global Communications Conference, Taipei, Taiwan, December. 2020 \cite{KimChoCuiLee:20}.} 
	\thanks{
		M.\ Kim and H.\ Cho are with the Department of Electrical Engineering and Computer Science,
		Daegu Gyeongbuk Institute of Science and Technology, Daegu, 42988, South Korea
		(e-mail:\texttt{\{alsrjs1807, nb00040\}@dgist.ac.kr}).}
	\thanks{
		Y.\ Cui is with the
		Department of Electronic Engineering, Shanghai Jiao
		Tong University, Shanghai 200240, China (e-mail: \texttt{cuiying@sjtu.edu.cn}).   
	}
	\thanks{J.\ Lee is with the Department of Electrical and Computer Engineering, Sungkyunkwan University (SKKU), Suwon 16419, Republic of Korea (e-mail: \texttt{jemin.lee@skku.edu}).}
}
\maketitle 
%

%

%
%
\acresetall
\begin{abstract}
In this paper, we consider the service caching and the computing resource allocation in \ac{EC} enabled networks. We introduce a random service caching design considering multiple types of latency sensitive services and the \acp{BS}' service caching storage. We then derive a \ac{SSP}. We also formulate a \ac{SSP} maximization problem subject to the service caching distribution and the computing resource allocation. Then, we show that the optimization problem is nonconvex and develop a novel algorithm to obtain the stationary point of the \ac{SSP} maximization problem by adopting the parallel \ac{SCA}. Moreover, to further reduce the computational complexity, we also provide a low complex algorithm that can obtain the near-optimal solution of the \ac{SSP} maximization problem in high computing capability region. Finally, from numerical simulations, we show that proposed solutions achieve higher \ac{SSP} than baseline schemes. Moreover, we show that the near-optimal solution achieves reliable performance in the high computing capability region. We also explore the impacts of target delays, a \acp{BS}' service cache size, and an \ac{EC} servers' computing capability on the \ac{SSP}.
\end{abstract}

\acresetall

\acresetall
\section{Introduction}
With the growing popularity of mobile applications and the IoT technology, the demand for computation intensive and latency sensitive services such as virtual reality is also increasing. It is hard for mobile users to meet the required deadline of the computation task, due to the limited battery capacity and the low computing capability of the mobile user. One promising solution for tackling this issue is the \ac{EC} technology, which utilizes the computing capability at the network edge such as the \acp{BS} \cite{MaoYouZha:17}. By applying the \ac{EC}, the computation task of a mobile user is offloaded to its nearby \ac{BS}. Once the computation is completed, the computation result is transmitted to the user, which results in lower latency compared to the local computing.

To reduce the cost of executing the task at the \ac{EC} server, the content caching has been considered in the \ac{EC}-enabled networks. In the content caching, each \ac{BS} caches the computation results of certain tasks frequently requested by users, so a \ac{BS} can directly transmit the computation results of the requested task to the user without executing the task at the \ac{EC} server again. Therefore, the execution cost at the \ac{EC} server can be reduced.

However, when users often use computation intensive applications including the cognitive assistance and virtual/augmented reality, different service software is required, and a \ac{BS} that caches certain service software can directly execute a task for the service. Therefore, to tackle these issues, service caching at the network edge, i.e., caching the software of services that are frequently requested by mobile users at \acp{BS}, has been recently introduced \cite{XuCheZho:18}. Supported by the service caching, a \ac{BS} that caches certain service software can directly compute a computation task that belongs to the certain service. Due to the limited storage resource, it is hard for a \ac{BS} to cache all service software. Therefore, it is critical to optimally select the service software to be cached at each \ac{BS}.

There have been several works on optimal communication and computing resource allocations and task offloading decisions in \ac{EC} enabled networks \cite{TraPom:18, LiuBenDebPoo:19, MaoZhaLet:16, QQMaoZhaLet:17}. For example, in \cite{TraPom:18, MaoZhaLet:16, LiuBenDebPoo:19}, the communication and computing resource allocations and task offloading decisions are jointly optimized to minimize the energy consumption and latency for computing and transmitting tasks. In \cite{QQMaoZhaLet:17}, the communication resource allocation and task offloading decisions are jointly optimized to minimize the energy consumption of computing and transmitting tasks. However, in \cite{TraPom:18, LiuBenDebPoo:19, MaoZhaLet:16, QQMaoZhaLet:17}, the authors assume that the numbers and locations of \acp{BS} and users are fixed, which makes their solutions ineffective when the network setup changes. Moreover, it can be unrealistic for a control tower to know all locations and channels of all \acp{BS} and users. Recently, randomly distributed \acp{BS} and users have been considered in several works which analyze the communication and computation latency in \ac{EC} enabled networks \cite{HuZonWanZhu:19, ParLee:20, KoHanHua:18}. Specifically, \cite{HuZonWanZhu:19,ParLee:20} consider multi tier heterogeneous networks and \cite{KoHanHua:18} studies single tier networks. Nevertheless, in the aforementioned works \cite{TraPom:18, LiuBenDebPoo:19, MaoZhaLet:16, QQMaoZhaLet:17,HuZonWanZhu:19,ParLee:20,KoHanHua:18}, the authors implicitly assume that all service software can be cached at each \ac{BS}. This assumption is unrealistic due to the limited storage resource at a \ac{BS}.

As tasks belonging to different services have possibly distinct computation workloads, service caching should be jointly optimized with the computing resource allocation. There have been some recent works on the joint optimization of the service caching, communication and computing resource allocations, and task offloading decisions in \ac{EC} enabled networks \cite{XuCheZho:18,TraChaPom:19,LiZhaJiLi:19,CheHaoHuHosGho:18,ZhaHouWanCha:18,WenCuiQueZheJin:20}. For example, in \cite{XuCheZho:18, TraChaPom:19, LiZhaJiLi:19, CheHaoHuHosGho:18}, the service caching and task offloading decisions are jointly optimized to minimize the energy consumption and latency for computing and transmitting tasks. In \cite{ZhaHouWanCha:18}, an online algorithm is presented for the dynamic service caching with arbitrary service request patterns. In \cite{WenCuiQueZheJin:20}, the authors optimize communication and computing resource allocations together with the service caching and task offloading decisions. However, in most works on service caching in \ac{EC} enabled networks such as \cite{XuCheZho:18, TraChaPom:19, ZhaHouWanCha:18, WenCuiQueZheJin:20, LiZhaJiLi:19, CheHaoHuHosGho:18}, fixed numbers of \acp{BS} and users with fixed locations are considered, which can be impractical for wireless networks of diverse deployments and mobile users with varying locations. On the other hand, some works such as \cite{CuiJia:16} consider random distributions of \acp{BS} and users for content caching by considering data intensive services rather than computation intensive services.

This motivates us to optimize the service caching and the computing resource allocation with randomly distributed \acp{BS} and users in the large-scale networks for randomly distributed BSs and users. Each user randomly requests a computation task that belongs to a particular service. Each \ac{BS} has an \ac{EC} server, and caches service software. We investigate two service time cases at the \ac{EC} server: the \ac{RT} and the \ac{DT} cases. Since BSs and users are randomly distributed in the networks, we first analyze the average networks performance, i.e., the \ac{SSP}, which is the probability of completing the service request, computation, and reception of the computation result are completed within their respective target delays. Then, we formulate the \ac{SSP} maximization problem subject to the service caching and the computing resource allocation and obtain the stationary point of the \ac{SSP} maximization problem.
The main contributions of this paper can be summarized as follows.

\begin{itemize}
	\item We newly provide the closed form expression of the approximated \ac{SSP} for both the \ac{RT} and the \ac{DT} cases using stochastic geometry and queuing theory. The approximated \ac{SSP} for the \ac{RT} case is a differential function, and the approximated \ac{SSP} for the \ac{DT} case is converted to a differential form using the Hausdorff approximation \cite{MarSyeKyuNikIliAntRahAse:18} for tractability in the later \ac{SSP} optimization.
	
	\item We jointly optimize the service caching and the computing resource allocation to maximize the \ac{SSP} for both the \ac{RT} and the \ac{DT} cases in the large-scale networks for randomly distributed \acp{BS} and users. To the best of our knowledge, this is the first time that jointly optimizes the service caching and computing resource allocation for randomly distributed \acp{BS} and users. 
	
	\item Since the formulated \ac{SSP} maximization problem is a challenging nonconvex problem with a large number of variables, we propose a parallel iterative algorithm to obtain a stationary point based on parallel \ac{SCA} \cite{RazMeiHonMinLuoZhiPanJon:14}. Note that the parallel \ac{SCA} yields lower computational complexity than the conventional \ac{SCA}.
	
	\item To further reduce the computational complexity, we also develop a low complex algorithm that obtains a near-optimal solution of the \ac{SSP} maximization problem for both the \ac{RT} and the \ac{DT} cases by using the asymptotically optimal solution obtained for the infinite computing capability case.
	
	\item From numerical results, we show the superiority of the proposed algorithms compared to baseline schemes. We also reveal the impacts of \acp{BS}' service cache size, computing capability of the \ac{EC} server, and target delays of each service on the \ac{SSP}. 
\end{itemize} 

The remainder of this paper is organized as follows. Section \ref{System model} describes the system model. Section \ref{Analysis} analyzes the \ac{SSP} for both the \ac{RT} and the \ac{DT} cases. Section \ref{Optimization} formulates the \ac{SSP} maximization problem in the \ac{EC} enabled network. Moreover, iterative algorithm to obtain the stationary point of  \ac{SSP} maximization problem is proposed. We then develop the iterative algorithm with low complexity to obtain the near-optimal solution of \ac{SSP} the maximization problem in Section \ref{Optimization_2}. Numerical results are provided in Section \ref{NumericalResults}. Finally, the conclusions are given in Section \ref{Conclusion}.

\acresetall

\section{System Model}\label{System model}
In this section, we present the system model of an \ac{EC}-enabled network with the randomly distributed \acp{BS} and users. Each user requests a computation task that belongs to service among multiple types of services, which can be computed by the service software cached at the \acp{BS}.
\subsection{Network Model}
\begin{figure}[t]
	\centering
	{
		\includegraphics[width=1\columnwidth]{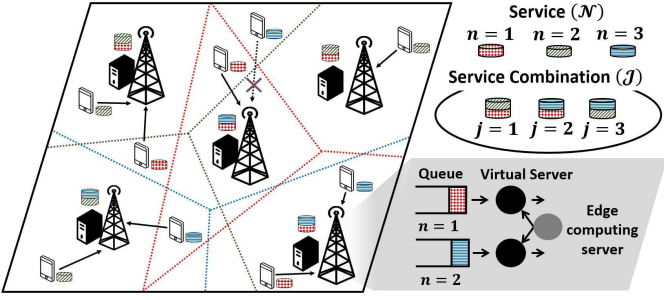}
		\vspace{-3mm}
		\caption{System model where \acp{BS}, equipped with a \ac{EC} server, have two different service software (i.e., $K=2$) among three different services (i.e., $N=3$). Each Voronoi cell (in the same color as the service) represents the \ac{BS} coverage for each service.}
		\label{fig: System model}
		\vspace{-3mm}
	}
\end{figure}

We consider an \ac{EC} enabled network which consists of single antenna \acp{BS} and single antenna users as shown in Fig. \ref{fig: System model}. The locations of \acp{BS} and users are modeled as independent homogeneous \acp{PPP} $\Appp{bs}$ and $\Appp{u}$, with spatial densities $\Adensity{bs}$ and $\Adensity{u}$, respectively. Each \ac{BS} is equipped with a cache and an \ac{EC} server. Therefore, each \ac{BS} has both caching and computing capabilities, while each user has neither caching nor computing capabilities\footnote{Note that the proposed algorithm in this paper can also be used for the case where not only \acp{BS} but also users have both the caching and computing capabilities. This can be done by adding the optimization parameters and the constraints of the user in a similar way to the ones for \acp{BS}.}. The computing capability of each \ac{BS} is denoted by $F_{\text{bs}}>0$ [number of CPU cycles per second], and the transmit power of \acp{BS} and users are denoted by $\Apower{bs}$ and $\Apower{u}$ [Watt], respectively. We consider both uplink and downlink transmissions operating in the \ac{TDD} mode. The channel bandwidth is $W$ [Hz], and each channel is reused in every $\kappa$ cell. As a consequence, the distribution of interfering \acp{BS} can be approximated into a \ac{PPP} $\tilde{\Phi}_{\text{bs}}$ with spatial density $\frac{\Adensity{bs}}{\kappa}$ \cite{ElsSulAloWin:17}. We consider both path loss and small scale fading. Due to path loss, a transmitted signal with distance $x$ is attenuated by a factor $x^{-\alpha}$, where $\alpha >2$ is the path loss exponent. For small scale fading, we assume Rayleigh fading channels.

We consider $N$ latency sensitive services, denoted by $\mathcal{N}\triangleq\{1,2,\ldots,N\}$. We assume each user randomly requests latency sensitive service $n \in \mathcal{N}$ with probability $p_n \in [0,1]$, where $\sum_{n \in \mathcal{N}} p_n = 1$. To provide service $n$ to the user, a certain computation task (i.e., task $n$) needs to be executed by its service software. Each task $n$ is characterized by three task parameters, i.e., the size of input data $\Tsize{i} > 0$ [bits], the size of computation result $\Tsize{o}>0$ [bits], and the size of computation workload $f_n$ [number of CPU cycles].

We consider a discrete time model, where time is divided into discrete slots. Specifically, in each time slot, each user randomly decides whether to request a service or not with probability $p_s \in [0,1]$. Once the user decides to request a service, we assume that the user randomly requests one service among $\mathcal{N}$. Here, we define $p_n$ as the probability that a user requests service $n \in \mathcal{N}$, where $\sum_{n\in\mathcal{N}} p_n=1$. Note that $p_n$ is the long-term average of users' service request, i.e., the service popularity in the large-scale EC-enabled networks. Regarding of the association, we consider the service centric association rule, where a user requesting a task $n$ associates with the nearest \ac{BS} caching the service software $n$. When user sends a service request to its serving \ac{BS}, the \ac{BS} executes the computation task of the requested service. After computing the task, the \ac{BS} transmits the computation result to the user. However, when the user is associated with the other \ac{BS} in present, the \ac{BS} transmits the computation result to the user through the other \ac{BS} associated with the user.

\subsection{Service Caching and Computation Model}\label{System-model}
In a practical \ac{EC} enabled network, each \ac{EC} server has limited computing capability and caching storage, so the only limited amount of service software can be cached \cite{XuCheZho:18}. Therefore, each \ac{BS} can cache the different amount of service software within its limit of caching storage. Let us define $\Omega_{\text{bs}}$ and $\Omega_n, n \in \mathcal{N}$ as the BSs' cache storage and the required storage size to cache the service software $n$, respectively. Then, we can obtain the set of possible service combinations $\mathcal{J}$ that satisfy the following constraints.
\begin{align}
	\sum_{k \in \mathcal{J}} \Omega_k \leq \Omega_{\text{bs}}. 
\end{align}
However, for analytical tractability, we consider that the same amount of service software is cached at each \ac{BS}. Then, without loss of generality, we assume that each \ac{BS} caches $K \left(\leq N \right)$ service software of the same size. Hence, there exist $J\triangleq\binom{N}{K}$ different service combinations that can be cached in a BS. Let $\mathcal{J}\triangleq\{1,2,\ldots,J\}$ denote the index set of all possible service combinations. 

We consider a \emph{random service caching} design, where a service combination is randomly adopted by each BS with a certain probability. The probability that service combination $j$ is cached in a BS is denoted by $a_j$, where
\begin{align}
&a_j \geq 0, \quad  j \in \mathcal{J}, \label{eq:C_1} \\
&\sum_{j \in \mathcal{J}} a_j = 1. \label{eq:C_2}
\end{align}
Here, we denote $\OptVar{a} \triangleq \left(a_j\right)_{j \in \mathcal{J}}$ as the service caching distribution. 
Furthermore, we define the set of service combinations which include service software $n$ as $\mathcal{J}_n$. From this, the probability that service software $n$ is cached at a \ac{BS}, denoted by $T_n$, where
\begin{align}
&T_n = \sum_{j \in \mathcal{J}_n} a_j, \quad  n \in \mathcal{N},
\label{eq:Relationship_a_T} \\
&0 \leq T_n \leq  1, \quad n \in \mathcal{N}, \label{eq:C1_special}\\
&\sum_{n \in \mathcal{N}} T_n = K. \label{eq:C2_special}
\end{align}
Here, we denote $\OptVar{T} \triangleq \left(T_n\right)_{n \in \mathcal{N}}$ as the service probability distribution. Note that the random service caching adopted in this paper is analogous to the random caching considered in \cite{CuiJia:16}.

We assume that each \ac{EC} server has $K$ virtual servers with different computing capabilities by applying hyper threading technology and asymmetric heterogeneous computing \cite{MarBinHilHinKouMilUpt:02}. Each virtual server has a queue with infinite capacity and can execute different service software cached in the \ac{BS}. In other words, we consider the \ac{EC} server consisting of $K$ queues and $K$ virtual servers, and each cached service is executed in a different set of the single queue and single virtual server. Regarding the service time at the server, we consider two cases: 1) \ac{RT} case and 2) \ac{DT} case. Specifically, the service time is modeled as an exponential random variable in the \ac{RT} case, and as constant time in the \ac{DT} case. Those two models are the ones, generally used in existing works\footnote{For example, the exponential distribution is used to model the service time at servers (e.g., edge and cloud-edge servers \cite{XiaAoShaSha:20,RomBen:20}, and the virtual machine \cite{QiaJeiXiaXia:20}), and proven as a realistic model from the measurements of video computation times at YouTube \cite{AbhSor:10}. The constant service time is also used for the edge computing system \cite{FedOlaBeaPet:21,ZinOsv:19} and the weather monitoring system \cite{QixJinLeiZhiZhuZhi:20}.}. The service time is denoted by $\ATimeQ{k}{Q}, k \in \{\text{m},\text{d}\}$, where $\text{`m'}$ and $\text{`d'}$ indicate the \ac{RT} and \ac{DT} cases, respectively. We adopt the \ac{FIFO} discipline at each queue and virtual server. The computing capability assigned for executing service software $n$ at the virtual server of the \ac{BS} which caches service combination $j$, is denoted by $b_{n,j}F_{\text{bs}}$, where
\begin{align}
&b_{n,j} \geq 0, \quad  j \in \mathcal{J}, \quad  n \in \mathcal{N}_j, \label{eq:C_3}\\
&\sum_{n \in \mathcal{N}_j} b_{n,j} = 1, \quad  j \in \mathcal{J}. \label{eq:C_4}
\end{align}
Here, $\mathcal{N}_j$ is the set of service software contained in combination $j$. We denote $\OptVar{b} \triangleq (b_{n,j})_{n \in \mathcal{N},j \in \mathcal{J}}$ as the computing resource allocation.
\subsection{Communication Model}
Each user first transmits the service request message to its serving \ac{BS}, corresponding to the uplink transmission. Without loss of generality, we assume that a typical user, denoted by $u_o$, is located at the origin, and the serving \ac{BS} of $u_o$ is denoted by $B_o^{(\text{U})}$. We consider the interference limited environment when users using the same frequency band as $u_o$ act as the interferer. Then, the maximum achievable uplink data rate for $B_o^{(\text{U})}$ at $u_o$, denoted by $\ARate{U}$, is given by
\begin{align}
\ARate{U} = \frac{W}{\kappa}\log_2\left(1+\frac{\left|h_{o,o}^{(\text{U})}\right|^2\left(d_{o,o}^{(\text{U})}\right)^{-\alpha}}{\sum_{l \in \Psi_{u_o}} \left|h_{l,o}^{(\text{U})}\right|^2\left(d_{l,o}^{(\text{U})}\right)^{-\alpha}}\right),
\end{align}
where $\Psi_{u_o}$ is the locations of interfering users \ac{w.r.t.} $u_o$. Here, $d_{l,t}^{(\text{U})}$ and $h_{l,t}^{(\text{U})}$ represent the distance and small scale fading channel gain of the link between $u_l$ and $B_t$, respectively. Moreover, the uplink transmission time of $u_o$'s service request $n$ is $\ATime{U} = \frac{\Tsize{i}}{\ARate{U}}$.

Once $B_o^{(\text{U})}$ finishes the computation of the service requested by $u_o$, it transmits the computation results to $u_o$ or the \ac{BS} associated with $u_o$ at present. We denote $B_o^{(\text{D})}$ as the \ac{BS} associated with $u_o$ at present. We assume that the transmission time between $B_o^{(\text{U})}$ and $B_o^{(\text{D})}$ is negligible due to the high capacity wired backhaul. Similar to the uplink scenario, \acp{BS} that use the same frequency band as $B_o^{(\text{D})}$ act as the interferer. Then, the maximum achievable data rate for $u_o$ at $B_o^{(\text{D})}$, denoted by $\ARate{D}$, is given by
\begin{align}
\ARate{D} = \frac{W}{\kappa}\log_2\left(1+\frac{\left|h_{o,o}^{(\text{D})}\right|^2\left(d_{o,o}^{(\text{D})}\right)^{-\alpha}}{\sum_{k \in \tilde{\Phi}_{\text{bs}}} \left|h_{o,k}^{(\text{D})} \right|^2\left(d_{o,k}^{(\text{D})}\right)^{-\alpha}} \right).
\end{align}
Then, the downlink transmission time of $u_o$'s computation result for service $n$ is $\ATime{D} = \frac{\Tsize{o}}{\ARate{D}}$.
\subsection{Performance Metric}
To satisfy that latency sensitive services require that the uplink transmission, computation, and downlink transmission are completed within their respective target delays as in \cite{XinZheYuaXin:21,ChaKaiHyuByo:16}, we introduce a \ac{SSP}, denoted by $\mathcal{P}(\OptVar{a},\OptVar{b})$, as the performance metric, given as\footnote{Note that our work can be easily extended to the case where the target delays are optimized for maximizing the \ac{SSP}.}
\begin{align}
\MetricSSP{a}{b} = \sum_{n \in \mathcal{N}} p_n \mathbb{P}&\left[\ATime{U} \leq \Tdead{U},\ATimeQ{k}{Q} \leq \Tdead{Q}, \right. \nonumber \\
&\left. \ATime{D} \leq \Tdead{D}\right], \, k \in \{\text{m},\text{d}   \},
\label{eq:SSP}
\end{align}
where $\Tdead{U}$, $\Tdead{Q}$, and $\Tdead{D}$ are the target delays for uplink transmission, computation, and downlink transmission, respectively. Here, $\ATime{U}$, $\ATimeQ{k}{Q}, k \in \{\text{m},\text{d}\}$, and $\ATime{D}$ can be correlated with each other, especially when the serving \ac{BS} in uplink and downlink transmission, i.e., $B_o^{(\text{U})}$, and $B_o^{(\text{D})}$ are the same. However, for analytical tractability, we assume that they are independent\footnote{Note that there exists a performance between among the SSPs in \eqref{eq:SSP} and \eqref{eq:SSP_ap} due to the independence assumption. However, in Section \ref{NumericalResults}, we show that the approximation error is small enough to be ignored.}, and hence, the \ac{SSP} in \eqref{eq:SSP} can be approximately expressed as
\begin{align}
\MetricSSPC{a}{b} =\sum_{n \in \mathcal{N}} &p_n \mathbb{P}\left[\ATime{U} \leq \Tdead{U}\right]\mathbb{P}\left[\ATimeQ{k}{Q} \leq \Tdead{Q}\right] \nonumber \\
&\times\mathbb{P}\left[\ATime{D} \leq \Tdead{D}\right] , \, k \in \{\text{m},\text{d} \}.
\label{eq:SSP_ap}
\end{align}
\section{Successful Service Probability Analysis}\label{Analysis}
In this section, we analyze the \ac{SSP} in both \ac{RT} and \ac{DT} cases. We first derive the \ac{SUTP}, the \ac{SDTP} and the \ac{SCPP}. We then present the \ac{SSP}. 
\subsection{Successful Uplink and Downlink Transmission Probabilities}
In this subsection, we derive the \ac{SUTP} and the \ac{SDTP} when a user requests a task $n$ and receives a computation result of the service $n$ from its associated \ac{BS}. For analytical tractability, we introduce the following assumption.
\begin{assumption}
	The location of uplink interfering users \ac{w.r.t.} the typical user follows the \ac{PPP}.
	\label{Ass:1}
\end{assumption}

Since one user exists in each cell and the user requests the task with probability $p_s$, the locations of interfering users are determined by thinning, dependent on \acp{BS}' locations. Therefore, the location of uplink interfering users does not follow a \ac{PPP}. However, according to \cite{NovDhiAnd:13}, it has been shown that this dependence is weak and can be negligible.

From Assumption \ref{Ass:1}, the distribution of uplink interfering users can be approximated to a \ac{PPP} with spatial density $\frac{p_s\Adensity{bs}}{\kappa}$. Therefore, the \ac{SUTP} of the user requesting a service $n$, denoted by $\Ametric{U}(\OptVar{a})$, is given by \cite{SinDhiAnd:13}
\begin{align}
	\Ametric{U}\left(\OptVar{a} \right) &= \mathbb{P}\left[\ARate{U} \geq \frac{\Tsize{i}}{\Tdead{U}} \right] \nonumber \\
	&=\frac{T_n}{T_n +\frac{p_s}{\kappa} Z\left(\TCoverage{U},\alpha,0\right)}, \quad n \in \mathcal{N},
	\label{eq:SUTP}
\end{align}
where $Z(a,b,c)\triangleq a^{2/b}\int_{(c/a)^{2/b}}^{\infty}\frac{1}{1+u^{b/2}}du$ and $\TCoverage{U} \triangleq 2^{\frac{\kappa\Tsize{i}}{ W\Tdead{U}}}-1$.

After computing the requested service of the user, the \ac{BS} transmits the computation result to the user. Hence, the \ac{SDTP} of a \ac{BS} transmitting the computation result of service $n$, denoted by $\Ametric{D}(\OptVar{a})$, is given by \cite{SinDhiAnd:13}
\begin{align}
	&\Ametric{D}\left(\OptVar{a} \right) = \mathbb{P}\left[\ARate{D} \geq \frac{\Tsize{o}}{\Tdead{D}} \right] \nonumber \\
	&\,=\frac{T_n}{T_n\left(1+\frac{p_s}{\kappa} Z\left(\TCoverage{D},\alpha,1\right)\right) \hspace{-1mm}+(1-T_n)\frac{p_s}{\kappa} Z\left(\TCoverage{D},\alpha,0\right)}, 
	\label{eq:SDTP}
\end{align}
for $n \in \mathcal{N}$, where $\TCoverage{D} \triangleq 2^{\frac{\kappa\Tsize{o}}{W\Tdead{D}}}-1$.

\subsection{Successful Computation Probability}
In this subsection, we derive the \ac{SCPP} in both the \ac{RT} and \ac{DT} cases. Since we adopt the service centric association rule, the task arrivals at each \ac{BS} is derived in the following proposition. 
\begin{proposition}
	The arrivals of task $n$ at a \ac{BS} that caches the service software $n$ can be modeled as a \ac{PP} $\Phi_{n}$ with an arrival rate $\lambda_{n}\left(\OptVar{a}\right)= \mathbb{E}\left[N_n \right] P_n^{(\text{U})}\left(\OptVar{a}\right)$.
\end{proposition}
\begin{IEEEproof}
	By applying the service centric association rule, the mean number of users requesting the task $n$ to a \ac{BS} is given by $\mathbb{E}[N_n]=1+ \frac{1.28p_np_{\text{s}}\Adensity{u}}{T_n\Adensity{bs}}$ \cite{SinDhiAnd:13}. Note that $\mathbb{E}[N_n]$ is independent of service combination $j$ cached at the \ac{BS}. This is because the mean number of users associated with the \ac{BS} is dependent on whether the \ac{BS} has service software $n$ or not. Moreover, each user requests a task independently, and the input data of task $n$ is transmitted successfully to the \ac{BS} with probability $\Ametric{U}\left(\OptVar{a} \right)$ in \eqref{eq:SUTP}. As a result, the task arrivals are modeled as \ac{PP} with an arrival rate $\lambda_{n}\left(\OptVar{a}\right)=\mathbb{E}[N_n]\Ametric{U}(\OptVar{a})$ \cite{KoHanHua:18}.
\end{IEEEproof}

When the input data of task $n$ arrives at the \ac{BS} that caches the service combination $j$, the service $n$ is scheduled at the \ac{BS}'s virtual server, and the computing capability $b_{n,j} F_{\text{bs}}$ is assigned for the computing service $n$. To guarantee the stability of the queue, we have the following constraint:
\begin{align}
	b_{n,j}\mu_n \geq \lambda_{n}\left(\OptVar{a}\right), \quad  n \in \mathcal{N}, \,  j \in \mathcal{J},
	\label{eq:stability_com}
\end{align}
where $\mu_n = \frac{F_{\text{bs}}}{f_n}$ is the maximum service rate at the virtual server for service $n$. Note that the condition in \eqref{eq:stability_com} guarantees that the arrival rate of task $n$ is smaller than the service rate of the virtual server assigned for service $n$. Then, we denote $\AmetricP{k}\left(\OptVar{a},\OptVar{b}\right), k=\{\text{m},\text{d}\}$ as the \ac{SCPP} of service $n$ at the \ac{BS} that caches service combination $j$, which is given by
\begin{align}
	\AmetricP{k}\left(\OptVar{a},\OptVar{b}\right) \hspace{-0.5mm}=\hspace{-0.5mm}\mathbb{P}\left[\hspace{-0.5mm}\ATimeQ{k}{Q} \leq \Tdead{Q} \left|\, j \in \mathcal{J}_n \right.\hspace{-0.5mm}\right]\hspace{-0.5mm}, n \in \mathcal{N}.
	\label{eq:SCPP_form}
\end{align} 
Now, from \eqref{eq:SCPP_form}, we derive the \ac{SCPP} of service $n$ for \ac{RT} and \ac{DT} cases.

\setcounter{eqnback}{\value{equation}}
\setcounter{equation}{21}
\begin{figure*}[t!]
	\normalsize
	\begin{align}
		\MetricSSPCC{m}{a}{b} =&\sum_{n \in \mathcal{N}} \frac{p_nT_n \sum_{j \in \mathcal{J}_n}a_j\left(1-e^{-b_{n,j}J_n+\lambda_{n}\left(\OptVar{a}\right) \Tdead{Q}} \right) }{D_nT_n^2+E_nT_n +A_nC_n}. \label{eq:SSP_close} \\
		\MetricSSPCC{d}{a}{b} = &\sum_{n \in \mathcal{N}} \frac{p_nT_n}{D_nT_n^2+E_nT_n +A_nC_n} \sum_{j\in \mathcal{J}_n} a_j\left(1-\frac{\lambda_{n}\left(\OptVar{a}\right)}{b_{n,j}\mu_n}\right) \sum_{\xi=0}^{\left\lfloor J_n\right\rfloor} \frac{\left(\lambda_{n}\left(\OptVar{a}\right)\right)^\xi\left(\frac{\xi}{b_{n,j}\mu_n}-\Tdead{Q}\right)^\xi e^{\lambda_{n}\left(\OptVar{a}\right)\left(\Tdead{Q}-\frac{\xi}{b_{n,j}\mu_n}\right)}}{\xi!\left(1+e^{-\delta\left(\xi-b_{n,j}J_n\right)}\right)e^{-\delta\left(\xi-b_{n,j}J_n\right)}}.
		\label{eq:SSP_close_2}
	\end{align}
	\vspace{-3mm}
	\hrulefill
\end{figure*}
\setcounter{equation}{\value{eqnback}}

\subsubsection{Random Service Time Case}
For the \ac{RT} case, the service time of the virtual server is modeled as an exponential random variable. Since the task arrivals follow \ac{PP}, we can model the computing process as the M/M/1 queue. Therefore, by applying the \ac{PDF} of sojourn time in M/M/1 queue \cite{Kle:75}, the \ac{SCPP} of service $n$ at the \ac{BS} that caches the service combination $j$ is given by
\begin{align}
	\AmetricP{\text{m}}\left(\OptVar{a},\OptVar{b}\right)= 1-e^{-(b_{n,j} \mu_n-\lambda_{n}\left(\OptVar{a}\right))\Tdead{Q}}, \quad n \in \mathcal{N}.
	\label{eq:SCPP}
\end{align}
\subsubsection{Deterministic Service Time Case}
For the \ac{DT} case, we can model the computing process as the M/D/1 queue with constant service time. Hence, by applying the \ac{CDF} of sojourn time in the M/D/1 queue \cite{FraJan:01}, the \ac{SCPP} of service $n$ at the \ac{BS} that caches the service combination $j$ is given by
\begin{align}
	\AmetricP{\text{d}}\left(\OptVar{a},\OptVar{b}\right) =&\sum_{\xi=0}^{\left\lfloor\Tdead{Q}\mu_n\right\rfloor} \frac{\mathds{1}\left(b_{n,j}<\ATimeQ{\text{d}}{Q}b_{n,j}\mu_n\right)\left(\lambda_{n}\left(\OptVar{a}\right)\right)^{\xi}}{\left(1-\frac{\lambda_{n}\left(\OptVar{a}\right)}{b_{n,j}\mu_n}\right)^{-1}} \nonumber \\
	&\times \frac{e^{\lambda_{n}\left(\OptVar{a}\right)\left(\Tdead{Q}-\frac{\xi}{b_{n,j}\mu_n}\right)}}{\left(\frac{\xi}{b_{n,j}\mu_n}-\Tdead{Q}\right)^{-\xi}\xi!},\, n \in \mathcal{N},
	\label{eq:SCPP_2}
\end{align}
where $\left\lfloor x \right\rfloor$ is the floor function which gives the greatest integer less than or equal to $x$ and $\mathds{1}\left(A\right)$ is the indicator function which gives the value $1$ if the condition $A$ is satisfied and the value $0$ otherwise. In \eqref{eq:SCPP_2}, the indicator function can be tightly bounded by the logistic sigmoid function using the Hausdorff approximation \cite{MarSyeKyuNikIliAntRahAse:18}, as follows:
\begin{align}
	\mathds{1}\left(b_{n,j}<\ATimeQ{\text{d}}{Q}b_{n,j}\mu_n\right) \approx \frac{e^{-\delta\left(\xi-\Tdead{Q}b_{n,j}\mu_n \right)}}{\left(1+e^{-\delta\left(\xi-\Tdead{Q}b_{n,j}\mu_n\right)}\right)},
	\label{eq:app_computing}
\end{align}
where $\delta$ is the cut radius of the sigmoid function. Note that the gap between the original indicator function and the logistic sigmoid function is very small for the large positive value of cut radius $\delta\gg0$, and can be ignored. Hence, the \ac{SCPP} in \eqref{eq:SCPP_2} can be approximated to a differentiable form such as $\tilde{P}_{n,j,\text{d}}^{(\text{Q})}\left(\OptVar{a},\OptVar{b}\right)$, given by
\begin{align}
	\tilde{P}_{n,j,\text{d}}^{(\text{Q})}\left(\OptVar{a},\OptVar{b}\right) =& \sum_{\xi=0}^{\left\lfloor\Tdead{Q}\mu_n\right\rfloor} \frac{\left(\hspace{-0.5mm}1-\frac{\lambda_{n}\left(\OptVar{a}\right)}{b_{n,j}\mu_n}\right)e^{\lambda_{n}\left(\OptVar{a}\right)\left(\Tdead{Q}-\frac{\xi}{b_{n,j}\mu_n}\right)}}{\left(\frac{\xi}{b_{n,j}\mu_n}-\Tdead{Q}\hspace{-0.5mm}\right)^{-\xi}\left(\lambda_{n}\left(\OptVar{a}\right)\right)^{-\xi}}  \nonumber \\
	&\times \frac{e^{-\delta\left(\xi-\Tdead{Q}b_{n,j}\mu_n \right)}}{\xi!\left(1+e^{-\delta\left(\xi-\Tdead{Q}b_{n,j}\mu_n\right)}\right)}, \, n \in \mathcal{N}.
	\label{eq:SCPP_2_ap} 
\end{align}
Later in Section VI, we will show that the approximation error is small enough to be ignored.

\subsection{Successful Service Probability}
The \ac{SSP} in \eqref{eq:SSP_ap} can be represented as
\begin{align}
	\MetricSSPC{a}{b} = &\sum_{n \in \mathcal{N}} p_n \Ametric{U}\left(\OptVar{a} \right)\Ametric{D}\left(\OptVar{a} \right)\sum_{j \in \mathcal{J}_n} \frac{a_j}{T_n}\tilde{P}_{n,j,k}^{(\text{Q})}\left(\OptVar{a},\OptVar{b}\right), \label{eq:SSP_sum} 
\end{align}
for $k \in \{\text{m},\text{d}  \}$, where $\Ametric{U}$ and $\Ametric{D}$ are in \eqref{eq:SUTP} and \eqref{eq:SDTP}, respectively. Here, $\tilde{P}_{n,j,k}^{(\text{Q})}\left(\OptVar{a},\OptVar{b}\right)$ is $\AmetricP{\text{m}}\left(\OptVar{a},\OptVar{b}\right)$ in \eqref{eq:SCPP} for $k=\text{m}$ and $\tilde{P}_{n,j,\text{d}}^{(\text{Q})}\left(\OptVar{a},\OptVar{b}\right)$ in \eqref{eq:SCPP_2_ap} for $k = \text{d}$.

By substituting \eqref{eq:SUTP}, \eqref{eq:SDTP}, \eqref{eq:SCPP}, and \eqref{eq:SCPP_2_ap} into \eqref{eq:SSP_sum}, the \acp{SSP} for the \ac{RT} and \ac{DT} cases are respectively given by \eqref{eq:SSP_close} and \eqref{eq:SSP_close_2}, as shown on the top at this page, where $ A_n = \frac{p_s}{\kappa} Z\left(\TCoverage{U},\alpha,0\right)$, $B_n = \frac{p_s}{\kappa} Z\left(\TCoverage{D},\alpha,1\right)$, $C_n = \frac{p_s}{\kappa} Z\left(\TCoverage{D},\alpha,0\right)$, $D_n = 1 + B_n - C_n$, $E_n =C_n + A_n\left(1 + B_n - C_n\right)$, and $J_n=\mu_n\Tdead{Q}$.

\section{Joint Optimization of Service Caching Distribution and Computing Resource Allocation}\label{Optimization}
In this section, we formulate the \ac{SSP} maximization problem, and then, by using the parallel \ac{SCA}, we propose an iterative algorithm to obtain the stationary point of the problem.
\subsection{Problem Formulation}
The \ac{SSP} is affected by the random service caching and the computing capability of the \ac{EC} server. Therefore, we maximize the \ac{SSP} by optimizing the service caching distribution $\OptVar{a}$ and the computing resource allocation $\OptVar{b}$ as formulated below. 
\begin{problem}[\ac{SSP} maximization]
	\begin{alignat}{2}
	&\max_{\OptVar{a}, \OptVar{b}}\quad  
	\MetricSSPC{a}{b}
	\nonumber \\  
	&\,\,\text{s.t.}\quad\,\,\,
	\eqref{eq:C_1},\eqref{eq:C_2},\eqref{eq:C_3},\eqref{eq:C_4},\eqref{eq:stability_com}, \nonumber
	\end{alignat} 
	\label{pro:Original}
\end{problem}
where $k \in \{\text{m},\text{d}\}$ and $\MetricSSPC{a}{b}$ is in \eqref{eq:SSP_close} and \eqref{eq:SSP_close_2}. Here, $T_n$ is the service caching probability, which is defined in \eqref{eq:Relationship_a_T}. Note that since \eqref{eq:C_1} and \eqref{eq:C_2} imply \eqref{eq:C1_special} and \eqref{eq:C2_special}, we can ignore \eqref{eq:C1_special} and \eqref{eq:C2_special}.

Then, we obtain an equivalent problem of Problem \ref{pro:Original} by including inequality constraints in \eqref{eq:stability_com} implicitly in the objective function using the logarithmic barrier term as follows.
\begin{problem}[Equivalent problem of Problem \ref{pro:Original}]
	\begin{alignat}{2}
	&\max_{\OptVar{a}, \OptVar{b}}\quad  
	\MetricSSPC{a}{b}  +\frac{1}{\omega}\phi\left(\OptVar{a},\OptVar{b}\right)
	\nonumber \\  
	&\,\,\text{s.t.}\quad\,\,\,
	\eqref{eq:C_1},\eqref{eq:C_2},\eqref{eq:C_3},\eqref{eq:C_4}, \nonumber
	\end{alignat} 
	\label{pro: Original_eq}
\end{problem} 
where $\omega\gg0$ is a large positive value and $\phi\left(\OptVar{a},\OptVar{b}\right) = \sum_{n \in \mathcal{N}}\sum_{j \in \mathcal{J}_n}\log\left(b_{n,j}\mu_n-\lambda_{n}\left(\OptVar{a}\right) \right)$. Note that Problem 2 becomes infeasible if the queue stability condition in \eqref{eq:stability_com} is not satisfied (i.e., $b_{n,j}\mu_n < \lambda_n\left(\textbf{a}\right)$) since the logarithm of a negative value is not defined.

In the \ac{RT} case $\left(\text{i.e.,\,\,} k=\text{m}\right)$, the objective function in Problem \ref{pro: Original_eq} is marginally concave \ac{w.r.t.} $\OptVar{b}$, but not jointly concave \ac{w.r.t.} $\OptVar{a}$ and $\OptVar{b}$. The constraints in Problem \ref{pro: Original_eq} are all linear. Therefore, Problem \ref{pro: Original_eq} for the \ac{RT} case is a nonconvex problem. On the other hand, in the \ac{DT} case $\left(\text{i.e.,\,\,} k=\text{d}\right)$, Problem \ref{pro: Original_eq} is also a nonconvex problem since the objective function is not jointly concave \ac{w.r.t.} $\mathbf{a}$ and $\mathbf{b}$ and the constraints are linear functions.
\subsection{Optimal Solution}
In general, it is hard to obtain a globally optimal solution for a nonconvex problem with an effective and efficient method. A classic goal for dealing with a nonconvex problem is to obtain a stationary point that satisfies the \ac{KKT} conditions. 
Hence, we propose an iterative algorithm to obtain the stationary point of Problem \ref{pro: Original_eq} by using the parallel \ac{SCA} \cite{RazMeiHonMinLuoZhiPanJon:14}. Specifically, we divide the variables $(\OptVar{a},\,\OptVar{b})$ into one block for $\OptVar{a}$ and $J$ blocks for $\OptVar{b}_j \triangleq (b_{n,j})_{n \in \mathcal{N}_j}$, since the constraints are block separable. At each iteration, we solve one problem \ac{w.r.t.} $\OptVar{a}$ and $J$ problems \ac{w.r.t.} $\OptVar{b}_j,  j \in \mathcal{J}$, in a parallel manner.
	\begin{itemize}
		\item For $\OptVar{a}$, we solve the approximate convex problem for both the \ac{RT} and the \ac{DT} cases.
		\begin{problem}[Approximate Problem \ref{pro: Original_eq} for $\OptVar{a}$ at iteration $r+1$]
			\begin{alignat}{2}
			\max_{\OptVar{a}}\quad \appsubAgen{k}, \quad \text{s.t.}\,\, \eqref{eq:C_1},\eqref{eq:C_2}, \,\,\, k \in \{\text{m},\text{d}\}, \nonumber 
			\end{alignat}
			\label{pro:Sub_A_gen}
		\end{problem} 
		where $\BIvar{a}$ and $\BIvar{b}$ are the service caching distribution, and computing resource allocation at iteration $r$, and $\appsubAgen{k}$ is the approximation of the objective function in Problem \ref{pro: Original_eq} for $\OptVar{a}$, which is obtained in \eqref{eq:approximation_f} later.
		\item For $\OptVar{b}_j, j \in \mathcal{J}$, the objective function in Problem \ref{pro: Original_eq} is separated into $J$ sub-objective functions, i.e.,
		\setcounter{equation}{23}
		\begin{align}
		\subBgen{k}\hspace{-1mm} =\hspace{-1mm}\sum_{n \in \mathcal{N}_j}&p_n \Ametric{U}\left(\OptVar{a} \right)\Ametric{D}\left(\OptVar{a} \right)\tilde{P}_{n,i,k}^{(\text{Q})}\left(\OptVar{a},\OptVar{b} \right) \nonumber \\
		&+\frac{\log\left(b_{n,j}\mu_n-\lambda_{n}\left(\OptVar{a}_k^{(r)}\right) \right)}{\omega},
		\label{eq:Iter_B_j}
		\end{align}
		for $k \in \{\text{m},\text{d}\}$. Since $\mathcal{F}_{\textbf{b}_j,k}\left(\textbf{b}_j,\textbf{a}_k^{(r)} \right)$ is negative semidefinite, the objective function in \eqref{eq:Iter_B_j} is a concave function with respect to $\textbf{b}_j$.
		
		Then, we solve the convex problem for the \ac{RT} case.
		\begin{problem}[Problem \ref{pro: Original_eq} for $\OptVar{b}_j$ at iteration $r+1$ in the \ac{RT} case]
			\begin{alignat}{2}
			\max_{\OptVar{b}_j}\quad \subBgen{\text{m}}, \quad \text{s.t.}\,\, \eqref{eq:C_3},\eqref{eq:C_4}. \nonumber 
			\end{alignat}
			\label{pro:Sub_B_gen_m}
		\end{problem}
		On the other hand, we solve the approximate convex problem for the \ac{DT} case.
		\begin{problem}[Approximate Problem \ref{pro: Original_eq} for $\OptVar{b}_j$ at iteration $r+1$ in the \ac{DT} case]
			\begin{alignat}{2}
			\max_{\OptVar{b}_j}\quad \appsubBgen{\text{d}}, \quad \text{s.t.}\,\, \eqref{eq:C_3},\eqref{eq:C_4}, \nonumber 
			\end{alignat}
			\label{pro:Sub_B_gen_d}
		\end{problem}
		where $\appsubBgen{\text{d}}$ is the approximation of $\subBgen{\text{d}}$ for $\OptVar{b}_j$, which is obtained in \eqref{eq:approximation_f_2} later.
	\end{itemize}
	Then, using the optimal solutions of Problem \ref{pro:Sub_A_gen}, \ref{pro:Sub_B_gen_m} and \ref{pro:Sub_B_gen_d}, we update the service caching distribution $\OptVar{a}$ and the computing resource allocation $\OptVar{b}$, respectively, in a parallel manner.

\subsubsection{Optimal Service Caching Distribution}
First, we choose the approximation function of $\MetricSSPC{a}{b}  -\frac{1}{\omega}\phi\left(\OptVar{a},\OptVar{b}\right), k \in \{\text{m},\text{d}\}$ by taking the second order Taylor expansion for $\OptVar{a}$, given by
\begin{align}
\appsubAgen{k} \triangleq&  -\OptVar{a}^{T}\OptVar{a} +\Bigg(2\BIvar{a}+\nabla_{\OptVar{a}} \Big(\MetricSSPC{a}{b}  \nonumber \\
&\left.\left.\left.-\frac{1}{\omega} \phi\left(\OptVar{a},\OptVar{b}\right)\right)\right|_{\OptVar{a}=\BIvar{a}}\right)^{T}\OptVar{a},
\label{eq:approximation_f}
\end{align}
where $\mathbf{x}^{T}$ is the transpose of $\mathbf{x}$. Note that the \ac{SCA} algorithm obtains the stationary point of Problem \ref{pro: Original_eq} to solve the convex problem \ref{pro:Sub_A_gen} using the approximated objective function in \eqref{eq:approximation_f}. Therefore, the optimal solution of Problem \ref{pro:Sub_A_gen} is given by
	\begin{align}
	\IOptsol{\OptVar{a}}{k} \triangleq \text{argmax}_{\OptVar{a} \in \mathcal{X}_{a}}\quad \appsubAgen{k}, \quad k \in \{\text{m},\text{d}\},
	\end{align}
	where $\mathcal{X}_a$ is the feasible convex set of constraints \eqref{eq:C_1} and \eqref{eq:C_2}.
Since Problem \ref{pro:Sub_A_gen} is the convex optimization problem, we can obtain $\IOptsol{\OptVar{a}}{k},\, k \in \{\text{m},\text{d} \}$ by using standard convex optimization techniques such as the interior point method \cite{BoyVan:B04}. Finally, we update $\OptVar{a}_k^{(r)}$ at iteration $r+1$ by
\begin{align}
\OptVar{a}_k^{(r+1)} = \BIvar{a} + \alpha^{(r+1)}\left(\IOptsol{\OptVar{a}}{k}- \BIvar{a}  \right), \quad k \in \{\text{m},\text{d}\},
\label{eq:UpdateB}
\end{align}
where $\alpha^{(r+1)}$ is a positive diminishing step size that satisfies
\begin{align}
&\alpha^{(r+1)} > 0,  \lim_{r\rightarrow\infty} \alpha^{(r+1)} = 0, \nonumber \\
&\sum_{r=1}^{\infty} \alpha^{(r+1)} =\infty, \lim_{r\rightarrow\infty} \left(\alpha^{(r+1)}\right)^{2} < \infty. 
\label{eq:convergence}
\end{align}

\subsubsection{Optimal Computing Resource Allocation}
In the \ac{RT} case, the optimal solution of Problem \ref{pro:Sub_B_gen_m} is given by
\begin{align}
\IOptsol{\OptVar{b}}{j,\text{m}} \triangleq \text{argmax}_{\OptVar{b}_j \in \mathcal{X}_{b}}\quad \subBgen{\text{m}},
\end{align}
where $\mathcal{X}_{b}$ is the feasible convex set of constraints \eqref{eq:C_3} and \eqref{eq:C_4}. Since Problem \ref{pro:Sub_B_gen_m} is convex and the strong duality holds, we can obtain $\IOptsol{\OptVar{b}}{j,\text{m}}$ using \ac{KKT} conditions.
\begin{lemma}[Optimal Solution of Problem \ref{pro:Sub_B_gen_m}]
	\begin{align}
	\bar{b}_{n,j,\text{m}}^{(r+1)}=\max\left(G_{n,j}^{-1}\left(\eta_j^{*(r+1)}\right),0\right), \, n \in \mathcal{N},\,j \in \mathcal{J}_n,
	\label{eq:op_server_splitting} 
	\end{align}
	where $\eta_j^{*(r+1)}$ satisfies $\sum_{n \in \mathcal{N}_j} \max\left(G_{n,j}^{-1}\left(\eta_j^{*(r+1)}\right),0\right) = 1$ and $G^{-1}\left(\cdot\right)$ is the inverse function of $G\left(\cdot\right)$, which is given by
	\begin{align}
	G_{n,j}(b_{n,j}) =& \frac{J_np_na_{j,\text{m}}^{(r)}T_{n,\text{m}}^{(r)} e^{\lambda_{n,\text{m}}^{(r)}\Tdead{Q}-b_{n,j}J_n}  }{D_n\left(T_{n,\text{m}}^{(r)}\right)^2+E_nT_{n,\text{m}}^{(r)}+A_n^2} \nonumber \\
	&+\frac{\mu_n}{\omega}e^{\phi_{n,j}\left(\mathbf{a}_{\text{m}}^{(r)},b_{n,j}\right)},
	\label{eq:Odp_b}
	\end{align}
	where $\phi_{n,j}\left(\mathbf{a}_{\text{m}}^{(r)},b_{n,j}\right)=\log\left(b_{n,j}\mu_n-\lambda_{n}\left(\mathbf{a}_{\text{m}}^{(r)}\right) \right)$.
	\label{lem:Op_b}
\end{lemma}
\begin{IEEEproof}
	See Appendix \ref{app:Op_b}.
\end{IEEEproof} 

In \eqref{eq:Odp_b}, $G(b_{n,j})$ is a decreasing function since its derivative is negative. As $G\left(\bar{b}_{n,j,\text{m}}^{(r+1)}\right)$ is monotonically decreasing with $\bar{b}_{n,j,\text{m}}^{(r+1)}$, $G^{-1}\left(\eta_j^{*(r+1)}\right)$ is also monotonically decreasing with $\eta_j^{*(r+1)}$. Therefore, $\eta_j^{*(r+1)}$ and $\bar{b}_{n,j,\text{m}}^{(r+1)}$ can be easily obtained by using the bisection method. Then, we update $\OptVar{b}_{j,\text{m}}^{(r)}$ at iteration $r+1$ by
\begin{align}
\OptVar{b}_{j,\text{m}}^{(r+1)} = \OptVar{b}_{j,\text{m}}^{(r)} + \alpha^{(r+1)}\left(\IOptsol{\OptVar{b}}{j,\text{m}} - \OptVar{b}_{j,\text{m}}^{(r)}  \right),\quad  j \in \mathcal{J}.
\label{eq:UpdateA}
\end{align}
In the \ac{DT} case, we choose the approximation function of $\subBgen{\text{d}}$ by taking the second order Taylor expansion for $\OptVar{b}_j$, which is given by
\begin{align}
&\appsubBgen{\text{d}} \nonumber \\
&\triangleq-\OptVar{b}_{j}^{T}\OptVar{b}_{j} +\left(2\OptVar{b}_{j,\text{d}}^{(r)}+\nabla_{\OptVar{b}_j} \subBgen{\text{d}} \hspace{-1mm} \left|_{\OptVar{b}_{j,\text{d}}=\OptVar{b}_{j,\text{d}}^{(r)}}\right.\hspace{-0.5mm}\right)^{T}\OptVar{b}_{j}, \nonumber \\  &\quad j \in \mathcal{J}.
\label{eq:approximation_f_2}
\end{align}

Then, the optimal solution of Problem \ref{pro:Sub_B_gen_d} is given by
\begin{align}
\IOptsol{\OptVar{b}}{j,\text{d}} \triangleq \text{argmax}_{\OptVar{b}_j \in \mathcal{X}_{b}}\quad \appsubBgen{\text{d}}.
\end{align}
Since Problem \ref{pro:Sub_B_gen_d} is the convex optimization problem, we can obtain $\IOptsol{\OptVar{b}}{j,\text{d}}$ by using the interior point method \cite{BoyVan:B04}. Then, we update $\OptVar{b}_{j,\text{d}}^{(r)}$ at iteration $r+1$ by
\begin{align}
\OptVar{b}_{j,\text{d}}^{(r+1)} = \OptVar{b}_{j,\text{d}}^{(r)} + \alpha^{(r+1)}\left(\IOptsol{\OptVar{b}}{j,\text{d}} - \OptVar{b}_{j,\text{d}}^{(r)}  \right),\quad  j \in \mathcal{J}.
\label{eq:UpdateA_2}
\end{align}

Finally, the details of the proposed iterative algorithm are summarized in Algorithm \ref{Al:1}. Based on Theorem 1 in \cite{RazMeiHonMinLuoZhiPanJon:14}, we can derive the following result in Lemma \ref{lem:convergence}.
\begin{algorithm}[t!]
	\caption{Obtaining A Stationary Point of Problem \ref{pro: Original_eq}}
	\label{Al:1}
	\begin{algorithmic}[1]
		\State Initialize $\OptVar{a}^{(0)},\OptVar{b}^{(0)}$ which are feasible solution of Problem \ref{pro: Original_eq}, and set $r=1$. 	
		\While{$\left|\left|\nabla \left(\MetricSSPC{a}{b}  -\frac{1}{\omega}\phi\left(\OptVar{a},\OptVar{b}\right)\right)\right|\right|^2 \geq \tau, \, k \in \{\text{m},\text{d}   \}$} 
		\State Obtain $\IOptsol{\OptVar{a}}{k}, \, k \in \{\text{m},\text{d}  \}$ by solving Problem \ref{pro:Sub_A_gen} using the interior point method.
		\State Update $\OptVar{a}_k^{(r)}, \, k \in \{\text{m},\text{d}  \}$ according to \eqref{eq:UpdateB}.
		\For{$j \in \mathcal{J}$}
		\If{$k = \text{m}$}
		\State Obtain $\IOptsol{\OptVar{b}}{j,\text{m}}$ according to \eqref{eq:op_server_splitting}.
		\State Update $\OptVar{b}_{j,\text{m}}^{(r)}$ according to \eqref{eq:UpdateA}.
		\Else
		\State Obtain $\IOptsol{\OptVar{b}}{j,\text{d}}$ by solving Problem \ref{pro:Sub_B_gen_d} using the interior point method.
		\State Update $\OptVar{b}_{j,\text{d}}^{(r)}$ according to \eqref{eq:UpdateA_2}.
		\EndIf
		\EndFor
		\State Set $r=r+1$.
		\EndWhile
	\end{algorithmic}
\end{algorithm}
\begin{lemma}[Convergence of Algorithm \ref{Al:1}]
	When the step size $\alpha^{(r+1)}$ satisfies the conditions in  \eqref{eq:convergence}, $\lim_{r \rightarrow \infty} \left(\OptVar{a}_k^{(r)},\OptVar{b}_k^{(r)}\right), \, k \in \{\text{m},\text{d} \}$ are always stationary points of Problem \ref{pro:Original}.
	\label{lem:convergence}
\end{lemma}
\begin{IEEEproof}
	First, it is obvious that $\appsubAgen{k}, k \in \{\text{m},\text{d}\}$, $\subBgen{\text{m}}$, and $\appsubBgen{\text{d}}$ are differentiable for any given $\OptVar{a}$ and $\OptVar{b}$, respectively. Moreover, the Hessian of $\appsubAgen{k}$ and $\appsubBgen{\text{d}}$ is always negative.
	Second, in \eqref{eq:approximation_f}, the gradient of $\appsubAgen{k}$ \ac{w.r.t.} $\mathbf{a}$ is the same as the gradient of objective function in Problem \ref{pro: Original_eq} \ac{w.r.t.} $\mathbf{a}$. Moreover, in \eqref{eq:approximation_f_2}, the gradient of $\appsubBgen{\text{d}}$ \ac{w.r.t.} $\mathbf{b}_j$ is the same as that of objective function in Problem \ref{pro: Original_eq} \ac{w.r.t.} $\mathbf{b}_j$.
	Finally, $\appsubAgen{k}$, $\subBgen{\text{m}}$, and $\appsubBgen{\text{d}}$ are smooth functions on the constraints in \eqref{eq:C_1},\eqref{eq:C_2},\eqref{eq:C_3}, and \eqref{eq:C_4}. Therefore, using Theorem 1 in \cite{RazMeiHonMinLuoZhiPanJon:14}, Lemma \ref{lem:convergence} is obtained.
\end{IEEEproof}

It is known that for a given step size and an initial point, the number of iterations in parallel \ac{SCA} is always constant \cite{RazMeiHonMinLuoZhiPanJon:14}. Thus, the complexity order of Algorithm \ref{Al:1} is identical to that of each iteration in it. At each iteration, we solve one problem \ac{w.r.t.} $\OptVar{a}$ and $J$ problems \ac{w.r.t.} $\OptVar{b}_j, j \in \mathcal{J}$. In the case of solving the problem \ac{w.r.t.} $\OptVar{a}$ using the interior point method, the computational complexity is $\mathcal{O}\left(N^{7K/2} \right)$. In the case of solving the problem \ac{w.r.t.} $\OptVar{b}_j,\,j \in \mathcal{J}$, we consider two cases for the computational complexity. In the \ac{RT} case, the bisection method is used, so the computation complexity is $\mathcal{O}\left(N^{K}K^{3} \right)$. On the other hand, in the \ac{DT} case, the interior point method is used, so the computational complexity is $\mathcal{O}\left(N^{K}K^{7/2} \right)$. Therefore, the overall complexity of Algorithm \ref{Al:1} is $\mathcal{O}\left(N^{7K/2}\right)$ regardless of the service time case.

Note that the service popularity, i.e., $p_n$, is the average value for certain time period and region. This means $p_n$ can vary over time and space. For that case, the proposed algorithm can be performed again to find the optimal solution for the changed environment\footnote{The frequency of updating the service caching and computing resource allocation can be the networks management issue.}.
\section{Asymptotic Solution for Joint Optimization of Service Caching Distribution and Computing Resource Allocation}\label{Optimization_2}
In Problem \ref{pro: Original_eq}, there exist $J$ optimization variables for $\OptVar{a}$ and $KJ$ optimization variables for $\OptVar{b}$. In addition, despite the use of parallel \ac{SCA}, obtaining the stationary point of Problem \ref{pro: Original_eq} requires the high complex algorithm, especially for a large $N$. Therefore, in this section, we consider an asymptotic version of Problem \ref{pro: Original_eq} for a special case that the computing capability of the \ac{EC} server is infinity, i.e., infinite computing capability case. Then, we develop an iterative algorithm to obtain an asymptotically optimal solution for this case. Finally, using the asymptotically optimal solution, we develop an low complex algorithm to obtain a near-optimal solution of Problem \ref{pro: Original_eq} in high computing capability region.

From equation \eqref{eq:SSP_close} and \eqref{eq:SSP_close_2}, we can see that the physical layer parameters $\left(\text{i.e.,}\,\, \alpha,W,\Adensity{u},\Adensity{bs},F_{\text{bs}}\right)$ and the design parameters $\left(\OptVar{a},\OptVar{b}\right)$ jointly affect the \ac{SSP}. Moreover, the impacts of physical layer parameters and the design parameters on $\MetricSSPC{a}{b}$ are coupled in a complex manner. Therefore, to make a low complex algorithm to maximize the \ac{SSP}, we need to simplify the \ac{SSP} to have decoupled effects from the physical layer and the design parameters. For this, we analyze the asymptotic \ac{SSP} in the infinite computing capability case. From \eqref{eq:SSP_close} and \eqref{eq:SSP_close_2}, it is obvious that the \ac{SSP} increases with the computing capability $F_{\text{bs}}$, so we have the following corollary.
\begin{corollary}[Asymptotic Performance]
	When $F_{\text{bs}} \rightarrow \infty$, for $k = \{\text{m},\text{d} \}$, we have 
	\begin{align}
	&\AMetricSSP{\OptVar{a}} \triangleq \lim_{F_{\text{bs}}\rightarrow \infty} \MetricSSPC{a}{b} \nonumber \\
	&\quad= \sum_{n \in \mathcal{N}} \frac{p_n\left(\sum_{j \in \mathcal{J}_n}a_j\right)^2}{D_n\left(\sum_{j \in \mathcal{J}_n}a_i\right)^2+E_n\left(\sum_{j \in \mathcal{J}_n}a_j\right) +A_nC_n}. \label{eq:AsynmtoticSSP}
	\end{align}
	\label{corollary_1}
\end{corollary}
\begin{IEEEproof}
	When $F_{\text{bs}} \rightarrow \infty$, we have $J_n = \frac{F_{\text{bs}}}{f_n}\Tdead{Q} \rightarrow \infty$. Thus, for $j \in \mathcal{J},\,  n \in \mathcal{N}_j$, we have
	\begin{align}
	&\exp\left({-b_{n,j}J_n+\lambda_{n}\Tdead{Q}}\right) \rightarrow 0,
	\label{eq:lim_service_rate} \\
	&\sum_{k=0}^{\left\lfloor J_n\right\rfloor} \frac{\lambda_{n}^k\left(\frac{k}{b_{n,j}\mu_n}-\Tdead{Q}\right)^ke^{\lambda_{n}\left(\Tdead{Q}-\frac{k}{b_{n,j}\mu_n}\right)-\delta\left(k-b_{n,j}J_n\right)}}{k!\left(1+e^{-\delta\left(k-b_{n,j}J_n\right)}\right)} \nonumber \\
	&\rightarrow 1.  \label{eq:lim_service_rate_2}
	\end{align}
	Therefore, by substituting \eqref{eq:lim_service_rate} and \eqref{eq:lim_service_rate_2} into \eqref{eq:SSP_close} and \eqref{eq:SSP_close_2}, respectively, we obtain $\AMetricSSP{\OptVar{a}}$, which is given by \eqref{eq:AsynmtoticSSP}.
\end{IEEEproof}

From Corollary \ref{corollary_1}, we can see that in infinite computing capability case, some physical layer parameters such as $\Adensity{u}$ and $\Adensity{bs}$, and the design parameter $\OptVar{b}$ do not affect the \ac{SSP}. Moreover, the impacts of the physical layer parameter $p_n$ and the design parameter $\mathbf{a}$ on \eqref{eq:AsynmtoticSSP} are more clearly shown. In addition, the asymptotic \ac{SSP} in \eqref{eq:AsynmtoticSSP} has a simpler form than $\MetricSSPCC{m}{a}{b}$ in \eqref{eq:SSP_close} and $\MetricSSPCC{d}{a}{b}$ in \eqref{eq:SSP_close_2}.

Now, we maximize the asymptotic \ac{SSP} in \eqref{eq:AsynmtoticSSP} by optimizing the service caching distirbution. Note that the stability condition of the queue in \eqref{eq:stability_com} is always satisfied due to the high computing capability. Therefore, we have the following asymptotic optimization problem.
\begin{problem}[Asymptotic \ac{SSP} maximization]
	\begin{alignat}{2}
	\bar{\OptVar{a}}_{\infty} \triangleq\,\, \arg&\max_{\OptVar{a}}\quad  
	\AMetricSSP{\OptVar{a}}
	\nonumber \\  
	&\quad\text{s.t.}\quad\,\,\, \nonumber \eqref{eq:C_1},\eqref{eq:C_2}. \nonumber
	\end{alignat} 
	\label{pro:Asymptotic Case}
\end{problem}
Note that $\bar{\OptVar{a}}_{\infty}$ is an asymptotically optimal solution of Problem \ref{pro:Original}. In Problem \ref{pro:Asymptotic Case}, by replacing $\AMetricSSP{\OptVar{a}}$ with $\AMetricSSP{\OptVar{T}}$, we have the following optimization problem.
\begin{problem}[Equivalent problem of Problem \ref{pro:Asymptotic Case}]
	\begin{alignat}{2}
	\bar{\mathbf{T}}_{\infty} \triangleq\,\, \arg&\max_{\OptVar{T}}\quad  
	\AMetricSSP{\OptVar{T}}
	\nonumber \\  
	&\quad\text{s.t.}\quad\,\,\, \nonumber \eqref{eq:C1_special},\eqref{eq:C2_special}. \nonumber
	\end{alignat} 
	\label{pro:Equivalent_Asymptotic Case}
\end{problem}
We can show that Problem \ref{pro:Equivalent_Asymptotic Case} is equivalent to Problem \ref{pro:Asymptotic Case}.
\begin{lemma}[Equivalence of Problem \ref{pro:Asymptotic Case} and \ref{pro:Equivalent_Asymptotic Case}]
	The solutions of Problem \ref{pro:Asymptotic Case} and Problem \ref{pro:Equivalent_Asymptotic Case} satisfy
	\begin{align}
	\sum_{j \in \mathcal{J}_n} \bar{a}_{j,\infty} = \bar{T}_{n,\infty}, \, n \in \mathcal{N}, \, \AMetricSSP{\bar{\OptVar{a}}_{\infty}}=\AMetricSSP{\bar{\OptVar{T}}_{\infty}}.
	\label{eq:condition of best asymptotic solution}
	\end{align}
	\label{relationship}
\end{lemma}
\begin{IEEEproof}
	For given $\bar{\OptVar{a}}_{\infty}$, we can easily obtain $\bar{\OptVar{T}}_{\infty}$ which satisfies the conditions in \eqref{eq:C1_special} and \eqref{eq:C2_special}. On the other hand, for given $\bar{\OptVar{T}}_{\infty}$, we can obtain 
		$\bar{\OptVar{a}}_{\infty}$ which satisfies the conditions in \eqref{eq:C_1} and \eqref{eq:C_2} by using the method in Fig. 1. of \cite{BlaGio:15}. Moreover, due to the communication performance is affected only by whether the \ac{BS} has the service software $n$ or not, for given any $\bar{\OptVar{a}}_{\infty}$ and $\bar{\OptVar{T}}_{\infty}$, the optimal value of Problem \ref{pro:Asymptotic Case} and Problem \ref{pro:Equivalent_Asymptotic Case} are the same, i.e., $\AMetricSSP{\bar{\OptVar{a}}_{\infty}}=\AMetricSSP{\bar{\OptVar{T}}_{\infty}}$.
\end{IEEEproof}

In Problem \ref{pro:Equivalent_Asymptotic Case}, the objective function is convex \ac{w.r.t.} $\OptVar{T}$ and all constraints are affine. Since Problem \ref{pro:Equivalent_Asymptotic Case} is a problem that maximizes the convex function, it is a nonconvex problem. Specifically, Problem \ref{pro:Equivalent_Asymptotic Case} is a special case of the \ac{DC} programming problem. 
\subsection{Asymptotic Optimal Solution For Infinite Computing Capability Case}
In this subsection, we provide an asymptotically optimal solution of Problem \ref{pro:Equivalent_Asymptotic Case} for infinite computing capability case. We propose an iterative algorithm to obtain the stationary point of Problem \ref{pro:Equivalent_Asymptotic Case} using the \ac{CCCP} \cite{ThiAnPha:18}. Specifically, at each iteration, we choose the approximation function of $\AMetricSSP{\OptVar{T}}$ by taking the linear approximation, given by
\begin{align}
\appsubT{\infty} \triangleq& \left(\nabla_{\OptVar{T}} \AMetricSSP{\OptVar{T}}|_{\OptVar{T}=\Ivar{T}}\right)^{T}\left(\OptVar{T}-\Ivar{T}\right) \nonumber \\
&+ \AMetricSSP{\Ivar{T}},
\label{eq:approximation_special}
\end{align}
where $\OptVar{T}^{(r)}$ is the service probability distribution at iteration $r$. Then, we update the service probability distribution $\OptVar{T}^{(r)}$ at iteration $r+1$ by
\begin{align}
\OptVar{T}^{(r+1)} = \text{argmax}_{\OptVar{T} \in \mathcal{X}_{t}} \appsubT{\infty},
\label{eq:Update_T}
\end{align}
where $\mathcal{X}_{t}$ is the feasible convex set of constraints \eqref{eq:C1_special} and \eqref{eq:C2_special}. Based on Theorem 4 in \cite{SriLan:09}, we can show the following result in Lemma \ref{lem:convergence_asymptotic}.
\begin{lemma}
	By using the update rule in \eqref{eq:Update_T}, the convergence point $\lim_{r \rightarrow \infty}\Ivar{T}$ is always a stationary point of Problem \ref{pro:Equivalent_Asymptotic Case}.
	\label{lem:convergence_asymptotic}
\end{lemma}
\vspace{-4mm}
\begin{IEEEproof}
	We can see that the objective function $\AMetricSSP{T}$ in Problem \ref{pro:Equivalent_Asymptotic Case} is a differentiable convex function. Moreover, we can see that the gradient of $\AMetricSSP{T}$ in Problem \ref{pro:Equivalent_Asymptotic Case} is continuous. By using the linear approximation function, we can always find an optimal solution $\mathbf{T}^{(r)}$ at each iteration, which satisfies the condition in \eqref{eq:Update_T}. Finally, the set of optimal solutions is compact on the constraints in \eqref{eq:C1_special} and \eqref{eq:C2_special}. Therefore, using Theorem 4 in \cite{SriLan:09}, Lemma \ref{lem:convergence_asymptotic} is obtained.
\end{IEEEproof}

It is known that for a given step size and an initial point, the number of iterations for \ac{CCCP} is constant \cite{ThiAnPha:18}. Thus, the computational complexity order of Problem \ref{pro:Equivalent_Asymptotic Case} is identical to that of each iteration in it. At each iteration, we solve one approximate convex problem \ac{w.r.t.} $\OptVar{T}$, so the computational complexity of solving the problem \ac{w.r.t.} $\OptVar{T}$ using the interior point method is $\mathcal{O}\left(N^{7/2} \right)$. Therefore, the overall complexity of solving Problem \ref{pro:Equivalent_Asymptotic Case} is $\mathcal{O}\left(N^{7/2}\right)$. 
\subsection{Near-Optimal Solution In High Computing Capability Region}
In this subsection, we consider the near-optimal solution in high computing capability region for given asymptotically optimal solution of the infinite computing capability case. Specifically, for given the asymptotically optimal solution of the service probability $\textbf{T}^*$ in \eqref{eq:condition of best asymptotic solution}, we formulate the SSP maximization problem, and develop an iterative algorithm with low computational complexity to obtain a near-optimal solution in the high computing capability region.

First of all, for given $\textbf{T}^*$, we now maximize the \ac{SSP} by optimizing the service caching distribution $\textbf{a}$ and the computing resource allocation $\textbf{b}$ in the high computing capability region.
\begin{problem}[\ac{SSP} maximization for given $\OptVar{T}^*$]
	\begin{alignat}{2}
	&\max_{\OptVar{a}, \OptVar{b}}\quad  
	\mathcal{P}_{k}(\OptVar{a},\OptVar{b},\OptVar{T}^*) 
	\nonumber \\  
	&\,\,\text{s.t.}\quad\,\,\,
	\eqref{eq:C_1},\eqref{eq:condition of best asymptotic solution},\eqref{eq:C_3},\eqref{eq:C_4},\eqref{eq:stability_com}, \nonumber
	\end{alignat} 
	\label{pro:Equivalent problem}
\end{problem}
where $k \in \{\text{m},\text{d}\}$. Here, $\mathcal{P}_{\text{m}}(\OptVar{a},\OptVar{b},\OptVar{T}^*)$ and $\mathcal{P}_{\text{d}}(\OptVar{a},\OptVar{b},\OptVar{T}^*)$ are obtained by substituting $\OptVar{T}^*$ into $\OptVar{T}$ in \eqref{eq:SSP_close} and \eqref{eq:SSP_close_2}, respectively. 

In the \ac{RT} case $\left(\text{i.e.,\,\,} k=\text{m}\right)$, the objective function in Problem \ref{pro:Equivalent problem} is a jointly concave function \ac{w.r.t.} $\OptVar{a}$ and $\OptVar{b}$. On the other hand, in the \ac{DT} case $\left(\text{i.e.,\,\,} k=\text{d}\right)$, the objective function in Problem \ref{pro:Equivalent problem} is an affine function \ac{w.r.t.} $\OptVar{a}$ but not jointly concave \ac{w.r.t.} $\OptVar{a}$ and $\OptVar{b}$. Therefore, for given $\OptVar{T}^*$, Problem \ref{pro:Equivalent problem} is a nonconvex problem.

To obtain the stationary point of Problem \ref{pro:Equivalent problem} that satisfies the \ac{KKT} condition, we develop an algorithm with low complexity for each service time case. 

\subsubsection{Random Service Time Case}
In the \ac{RT} case, we divide Problem \ref{pro:Equivalent problem} into one master problem \ac{w.r.t.} $\OptVar{a}$ and $J$ subproblems \ac{w.r.t.} $\OptVar{b}_j$ by noting that the constraints on $J+1$ problems are separable. We first obtain an optimal solution of $J$ subproblems \ac{w.r.t.} $\OptVar{b}_j$. Then, for a given optimal solution of the subproblems \ac{w.r.t.} $\OptVar{b}_j$, we obtain the optimal solution of the master problem \ac{w.r.t.} $\OptVar{a}$. The details are further illustrated below. The master problem is formulated as follows.
\begin{problem}[Master problem - service caching distribution]
	\begin{alignat}{2}
	& \max_{\OptVar{a}}\quad  
	\sum_{n \in \mathcal{N}} \AMetricSSP{\OptVar{T}^*} + \sum_{j \in \mathcal{J}} \OptsubAspeR{m}a_j
	\nonumber \\  
	&\,\,\text{s.t.}\quad\,\,\,
	\eqref{eq:C_1},\eqref{eq:condition of best asymptotic solution}, \nonumber
	\end{alignat} 
	\label{pro:master-problem}
\end{problem}
where $\OptsubAspeR{m}$ is given by the following subproblem.
\begin{problem}[Subproblem - computing resource allocation]
	\begin{alignat}{2}
	\OptsubAspeR{m} \triangleq & \max_{\OptVar{b}_j}\quad  
	\subAspeR{m}
	\nonumber \\  
	&\,\,\text{s.t.}\quad\,\,\,
	\eqref{eq:C_3},\eqref{eq:C_4},\eqref{eq:stability_com}, \nonumber
	\end{alignat} 
	\label{pro:sub-problem}
\end{problem} 
where $\subAspeR{m}$ is given by
\begin{align}
\subAspeR{m} =-\sum_{n \in \mathcal{N}_j}&\frac{p_n}{T_n^*} \Ametric{U}\left(\OptVar{T}^* \right)\Ametric{D}\left(\OptVar{T}^* \right) \nonumber \\
&\times e^{-(b_{n,j} \mu_n-\lambda_{n}^{*})\Tdead{Q}},
\end{align}
where $\lambda_{n}^{*}=\frac{T_n^*+C_n}{T_n^*+A_n}$.

Since Problem \ref{pro:sub-problem} is convex and the strong duality holds, we obtain the optimal solution $\bar{\OptVar{b}}_{j,\text{m}}$ of Problem \ref{pro:sub-problem} by using \ac{KKT} conditions.
\begin{lemma}[Optimal Solution of Problem \ref{pro:sub-problem}]
	\begin{align}
	\bar{b}_{n,j,\text{m}} = \frac{1}{J_n}&\left\{\log\left(\frac{J_np_n(T_n^*)^2}{\hat{\zeta_j}^{*}\left(D_n(T_n^*)^2+E_nT_n^*+A_n^2\right)} \right)\right.\nonumber \\
	&\,\,\,\,+\lambda_{n}^{*}\Tdead{Q}\Bigg\},\,n \in \mathcal{N}_j,\,j \in \mathcal{J},
	\label{eq:op_special_b}
	\end{align}
	where $\hat{\zeta_j}^{*}$ is the Lagrange multiplier which satisfies the constraint in \eqref{eq:C_4}.
	\label{lemma: op_sub_problem}
\end{lemma}
\begin{IEEEproof}
	See Appendix \ref{app:op_sub_problem}.
\end{IEEEproof}

Next, we solve Problem \ref{pro:master-problem}. From Lemma \ref{lemma: op_sub_problem}, for given $\bar{\OptVar{b}}_{\text{m}}$, Problem \ref{pro:master-problem} becomes:
\begin{problem}[Service caching distribution for given $\bar{\OptVar{b}}_{\text{m}}$]
	\begin{alignat}{2}
	&\max_{\OptVar{a}}\quad  
	\sum_{n \in \mathcal{N}} \AMetricSSP{\OptVar{T}^*}-\sum_{j \in \mathcal{J}} \sum_{n \in \mathcal{N}_j} \frac{\hat{\zeta}_j^*}{J_n}a_j
	\nonumber \\  
	&\,\,\text{s.t.}\quad\,\,\,
	\eqref{eq:C_1},\eqref{eq:condition of best asymptotic solution}. \nonumber
	\end{alignat} 
	\label{pro:master-problem_2}
\end{problem}

In Problem \ref{pro:master-problem_2}, the objective function and all constraints are affine functions, so it is a \ac{LP} problem. To reduce the complexity, we formulate a dual problem of Problem \ref{pro:master-problem_2} with the dual variable $\mathbf{w}=(w_n)_{n\in\mathcal{N}}$. When $\bar{\OptVar{a}}_{\text{m}}$ is the optimal solution of Problem \ref{pro:master-problem_2}, we have
\begin{align}
\sum_{j \in \mathcal{J}}\sum_{n \in \mathcal{N}_j} -\frac{\zeta_j^*}{J_n}\bar{a}_{j,\text{m}} = \sum_{n \in \mathcal{N}}\bar{w}_nT_n^*,
\label{eq:dual_1}
\end{align}
where $\bar{\OptVar{w}}$ is the optimal solution of the dual problem, which is formulated as follows.

\begin{problem}[Dual problem of Problem \ref{pro:master-problem_2}]
	\begin{alignat}{2}
	\bar{\OptVar{w}} \triangleq\,\, \arg&\max_{\OptVar{w}}\quad  
	\sum_{n \in \mathcal{N}} w_nT_n^*
	\nonumber \\  
	&\,\,\text{s.t.}\quad\,\,\,
	\sum_{n \in \mathcal{N}_j}w_n\leq \sum_{n \in \mathcal{N}_j}-\frac{\zeta_j^*}{J_n}, \quad j \in \mathcal{J}. \nonumber
	\end{alignat} 
	\label{pro:dual_master_problem}
\end{problem}

Since Problem \ref{pro:dual_master_problem} is a convex problem, we can easily obtain an optimal solution $\bar{\OptVar{w}}$ by using the interior point method \cite{BoyVan:B04}. Then, from \eqref{eq:dual_1}, for given $\bar{\OptVar{w}}$, we can easily obtain the optimal solution $\bar{\OptVar{a}}_{\text{m}}$. Finally, we can obtain the near-optimal solution $\bar{\OptVar{a}}_{\text{m}}$ and $\bar{\OptVar{b}}_{\text{m}}$ of Problem \ref{pro:Equivalent problem}, as summarized in Algorithm \ref{Al:3}.

Now, we consider the computational complexity. In the \ac{RT} case, since the optimal solution of Problem \ref{pro:sub-problem} is expressed as the closed form in \eqref{eq:op_special_b}, the computational complexity of solving the problem \ac{w.r.t.} $\OptVar{b}_j$ is $\mathcal{O}\left(K \right)$. Hence, the computational complexity of solving the problem \ac{w.r.t.} $\OptVar{b}$ is $\mathcal{O}\left(N^{K}K \right)$, where $N^{K}$ is the number of service combinations. In Problem \ref{pro:dual_master_problem}, the interior point method is used, so the total computational complexity is $\mathcal{O}\left(N^{2+3K/2} \right)$. We can argue that the computational complexity in obtaining the near-optimal solution of Problem \ref{pro:Equivalent problem} is lower than that of obtaining the optimal solution of Problem \ref{pro: Original_eq}, which is $\mathcal{O}\left(N^{7K/2}\right)$.

\subsubsection{Deterministic Service Time Case}
In the \ac{DT} case, we propose an iterative algorithm to obtain the stationary point of Problem \ref{pro:Equivalent problem} by using parallel \ac{SCA}. Specifically, we divide the variables $(\OptVar{a},\,\OptVar{b})$ into one block for $\OptVar{a}$ and $J$ blocks for $\OptVar{b}_j$ since the constraints are block separable. At each iteration, we first solve one convex problem \ac{w.r.t.} $\OptVar{a}$. 
\begin{problem}[Approximate Problem \ref{pro:Equivalent problem} for $\OptVar{a}$ at iteration $r+1$]
	\begin{alignat}{2}
	&\max_{\OptVar{a}}\quad  
	\mathcal{P}_{\text{d}}\left(\OptVar{a},\hat{\OptVar{b}}_{\text{d}}^{(r)},\OptVar{T}^*\right) 
	\nonumber \\  
	&\,\,\text{s.t.}\quad\,\,\,
	\eqref{eq:C_1},\eqref{eq:condition of best asymptotic solution}. \nonumber
	\end{alignat} 
	\label{pro:Convex_equivalent_problem}
\end{problem}
where $\hat{\OptVar{b}}_{\text{d}}^{(r)}$ is the computing resource allocation for given $\mathbf{T}^*$ at iteration $r$, which is obtained in Problem \ref{pro:ConddvexP_2} later. The objective function of Problem \ref{pro:Convex_equivalent_problem} is an affine function \ac{w.r.t.} $\OptVar{a}$ and all constraints are affine functions, so Problem \ref{pro:Convex_equivalent_problem} is a \ac{LP} problem. To reduce the computational complexity, we formulate the dual problem of Problem \ref{pro:Convex_equivalent_problem} with dual variables $\mathbf{v}=(v_n)_{n \in \mathcal{N}}$. Specifically, when $\bar{\OptVar{a}}_{\text{d},\OptVar{T}^*}^{(r+1)}$ is the optimal solution of Problem \ref{pro:Convex_equivalent_problem}, we have
\begin{align}
\mathcal{P}_{\text{d}}\left(\bar{\OptVar{a}},\hat{\OptVar{b}}_{\text{d}}^{(r)},\OptVar{T}^*\right) = \sum_{n \in \mathcal{N}}\bar{v}_n^{(r+1)}T_n^*.
\label{eq:dual_2}
\end{align}
where $\bar{\mathbf{v}}^{(r+1)}$ is the optimal solution of dual problem, which is formulated as follows.
\begin{problem}[Dual problem of Problem \ref{pro:Convex_equivalent_problem}]
	\begin{alignat}{2}
	\bar{\mathbf{v}}^{(r+1)} \triangleq \,\,&\max_{\OptVar{v}}\quad  
	\sum_{n \in \mathcal{N}} v_nT_n^*
	\nonumber \\  
	&\,\,\text{s.t.}\quad\,\,\,
	\sum_{n \in \mathcal{N}_j} v_n\leq G_j\left(\hat{\mathbf{b}}_\text{d}^{(r)},\mathbf{T}^*\right), \quad j \in \mathcal{J}, \nonumber
	\end{alignat} 
	\label{pro:dual_master_problem_2}
\end{problem}
where $G_j\left(\hat{\mathbf{b}}_\text{d}^{(r)},\mathbf{T}^*\right), j \in \mathcal{J}$ is given by
\begin{align}
&G_j\hspace{-0.8mm}\left(\hat{\mathbf{b}}_\text{d}^{(r)},\mathbf{T}^*\hspace{-0.5mm}\right)\hspace{-1mm} = \hspace{-1.2mm}\sum_{n \in \mathcal{N}_j}\hspace{-1mm}\frac{p_n}{T_n^*} \Ametric{U}\left(\OptVar{T}^* \right)\Ametric{D}\left(\OptVar{T}^* \right)\hspace{-1mm}\left(1-\frac{\lambda_n^*}{\hat{b}_{n,j,\text{d}}^{(r)}\mu_n} \right) \nonumber \\
&\,\,\,\,\quad\times\sum_{k=0}^{\left\lfloor J_n\right\rfloor} \frac{\left(\lambda_n^*\right)^k\left(\frac{k}{\hat{b}_{n,j,\text{d}}^{(r)}\mu_n}-\Tdead{Q}\right)^ke^{\lambda_{n}^*\left(\Tdead{Q}-\frac{k}{\hat{b}_{n,j,\text{d}}^{(r)}\mu_n}\right)}}{k!\left(1+e^{-\delta\left(k-\hat{b}_{n,j,\text{d}}^{(r)}J_n\right)}\right)e^{-\delta\left(k-\hat{b}_{n,j,\text{d}}^{(r)}J_n\right)}}.
\end{align}

Since Problem \ref{pro:dual_master_problem_2} is a convex problem, we obtain an optimal solution $\bar{\mathbf{v}}^{(r+1)}$ by using the interior point method \cite{BoyVan:B04}. Therefore, from \eqref{eq:dual_2}, we can easily obtain the optimal solution $\bar{\OptVar{a}}_{\text{d},\OptVar{T}^*}^{(r+1)}$ for given $\bar{\mathbf{v}}^{(r+1)}$. Then, we update the service caching distribution at iteration $r+1$, $\bar{\OptVar{a}}_{\text{d}}^{(r)}$, by
\begin{align}
\hat{\OptVar{a}}_{\text{d}}^{(r+1)} = \hat{\OptVar{a}}_{\text{d}}^{(r)} + \alpha^{(r+1)}\left(\bar{\OptVar{a}}_{\text{d},\OptVar{T}^*}^{(r+1)} - \hat{\OptVar{a}}_{\text{d}}^{(r)}  \right).
\label{eq:Update_equivalent}
\end{align}

\begin{algorithm}[t!]
	\caption{Obtaining a near-optimal solution of Problem \ref{pro: Original_eq}}
	\label{Al:3}
	\begin{algorithmic}[1]
		\State Initialize $\OptVar{T}^{0}$ which is feasible to Problem \ref{pro:Equivalent_Asymptotic Case}, and set $r=1$.
		\While{$\left|\left|\nabla \AMetricSSP{\Ivar{T}}\right|\right|^2 \geq \tau$} 
		\State Obtain $\Ivar{T}$ according to \eqref{eq:Update_T}.
		\State Set $r = r+1$.
		\EndWhile
		\State Obtain $\bar{a}_{j,k}, j \in \mathcal{J}', k \in \{\text{m},\text{d} \}$ according to \eqref{eq:Relationship_best_asymptotic}.
		\If{$k = \text{m}$}
		\State Obtain the near-optimal solution $\bar{\OptVar{b}}_{\text{m}}$ according to \eqref{eq:op_special_b}.
		\State Obtain $\bar{a}_{i,\text{m}}, j \in \mathcal{J}\backslash
		\mathcal{J}'$ by solving \ref{pro:master-problem_2} using the interior point method.
		\Else
		\State Initialize $\OptVar{a}_{\text{d}}^{0},\OptVar{b}_{\text{d}}^{0}$ which are feasible to Problem \ref{pro:Equivalent problem}, and set $r=1$.
		\While{$||\nabla \mathcal{P}_{\text{d}}(\OptVar{a},\OptVar{b},\OptVar{T}^*)||^2 \geq \tau$}
		\State Obtain $\bar{a}_{j,\text{d},\OptVar{T}^*}^{(r+1)}, j \in \mathcal{J}\backslash\mathcal{J}'$ by solving Problem \ref{pro:Convex_equivalent_problem} using the interior point method.
		\State Update $\hat{\OptVar{a}}_{\text{d}}^{(r)}$ according to \eqref{eq:Update_equivalent}.
		\For{$j \in \mathcal{J}$}
		\State Obtain $\bar{\OptVar{b}}_{j,\text{d},\OptVar{T}^*}^{(r+1)}$ by solving Problem \ref{pro:ConddvexP_2} using the interior point method.
		\State Update $\hat{\OptVar{b}}_{j,\text{d}}^{(r)}$ according to \ref{eq:UpdateddA_2}.
		\EndFor
		\State Set $r = r+1$.
		\EndWhile
		\EndIf
	\end{algorithmic}
\end{algorithm}

Now, we solve the $J$ approximate convex problems \ac{w.r.t.} $\OptVar{b}_j,  j \in \mathcal{J}$. The objective function in Problem \ref{pro:Equivalent problem} is separated into $J$ sub-objective functions, i.e.,
\begin{align}
\mathcal{H}_{j,\text{d}}\left(\mathbf{b}_j,\hat{a}_{j,\text{d}}^{(r)}, \mathbf{T}^{*}\right) =\sum_{n \in \mathcal{N}_j}&p_n \Ametric{U}\left(\OptVar{T}^* \right)\Ametric{D}\left(\OptVar{T}^* \right) \nonumber \\
&\times \tilde{P}_{n,j,\text{m}}^{(\text{Q})}\left(\OptVar{b}_j,\hat{a}_{j,\text{d}}^{(r)} \right), \, j \in \mathcal{J}.
\end{align}

Then, for each $j \in \mathcal{J}$, we choose the approximation function of $\mathcal{H}_{j,\text{d}}\left(\mathbf{b}_j,\hat{a}_{j,\text{d}}^{(r)}, \mathbf{T}^{*}\right)$ by taking the second order Taylor expansion for $\OptVar{b}_j$, which is given by
\begin{align}
&\tilde{\mathcal{H}}_{j,\text{d}}\left(\mathbf{b}_j,\hat{a}_{j,\text{d}}^{(r)},\hat{\mathbf{b}}_{j,\text{d}}^{(r)} ,\mathbf{T}^{*}\right) \triangleq -\OptVar{b}_{j}^{T}\OptVar{b}_{j} -\left(-2\hat{\OptVar{b}}_{j,\text{d}}^{(r)} \right.\nonumber \\
&\quad\left.+\nabla_{\OptVar{b}_j} \mathcal{H}_{j,\text{d}}\left(\mathbf{b}_j,\hat{a}_{j,\text{d}}^{(r)}, \mathbf{T}^{*}\right)\left|_{\OptVar{b}_{j,\text{d}}=\hat{\OptVar{b}}_{j,\text{d}}^{(r)}}\right.\right)^{T}\OptVar{b}_{j}.
\label{eq:approximation_f_2dd}
\end{align}
Then, we formulate the following approximate convex problem \ac{w.r.t.} $\OptVar{b}_j,  j \in \mathcal{J}$.
\begin{problem}[Approximate Problem \ref{pro:Equivalent problem} for $\OptVar{b}_j$ at iteration $r+1$]
	\begin{alignat}{2}
	&\max_{\OptVar{b}_j}\quad  \tilde{\mathcal{H}}_{j,\text{d}}\left(\mathbf{b}_j,\hat{a}_{j,\text{d}}^{(r)},\hat{\mathbf{b}}_{j,\text{d}}^{(r)} ,\mathbf{T}^{*}\right) \nonumber \\  
	&\,\,\text{s.t.}\quad\,\,\, \nonumber \eqref{eq:C_3},\eqref{eq:C_4},\eqref{eq:stability_com}. \nonumber
	\end{alignat}
	\label{pro:ConddvexP_2}
\end{problem}
Since Problem \ref{pro:ConddvexP_2} is a convex problem, we can obtain an optimal solution $\bar{\OptVar{b}}_{j,\text{d}}^{(r+1)}$ by using the interior point method \cite{BoyVan:B04}. Then, we update $\hat{\mathbf{b}}_{j,\text{d}}^{(r)}$ at iteration $r+1$ by
\begin{align}
\hat{\mathbf{b}}_{j,\text{d}}^{(r+1)} = \hat{\mathbf{b}}_{j,\text{d}}^{(r)} + \alpha^{(r+1)}\left(\bar{\OptVar{b}}_{j,\text{d}}^{(r+1)} - \hat{\mathbf{b}}_{i,\text{d}}^{(r)}  \right),\quad  j \in \mathcal{J}.
\label{eq:UpdateddA_2}
\end{align}

In the \ac{DT} case, since the number of iterations for parallel \ac{SCA} is constant \cite{RazMeiHonMinLuoZhiPanJon:14}, the complexity order is identical to that of each iteration in it. At each iteration, we solve one convex problem \ac{w.r.t.} a and $J$ approximate convex problems \ac{w.r.t.} $b_j, j \in \mathcal{J}$, respectively. The computational complexity of solving the problem \ac{w.r.t.} $\OptVar{a}$ using the interior point method is $\mathcal{O}\left(N^{2+3K/2} \right)$. On the other hand, the computational complexity of solving the problem \ac{w.r.t.} $\OptVar{b}_i$ using the interior point method is $\mathcal{O}\left(N^{K}K^{7/2}\right)$. Finally, the total computational complexity is $\mathcal{O}\left(N^{2+3K/2} \right)$. We can argue that the computational complexity of obtaining the near-optimal solution of Problem \ref{pro:Equivalent problem} is lower than that of obtaining the optimal solution of Problem \ref{pro: Original_eq}. 

Finally, we can obtain the near-optimal solution of Problem \ref{pro:Equivalent problem}, as summarized in Algorithm \ref{Al:3}. Note that some service combination probabilities are obviously determined from Corollary \ref{corollary_1}, specifically, $a_j = 0, j \in \mathcal{J}_n$ if $T_n =0$ and $a_j = 0, j \notin \mathcal{J}_n$ if $T_n =1$. Therefore, to reduce the computational complexity, we first obtain the near-optimal solution of service combination probability from Corollary \ref{corollary_1} before solving the optimization problem.
\begin{align}
\bar{a}_{j,k} = 0, \quad j \in \mathcal{J}', \,\, k \in \{\text{m},\text{d} \},
\label{eq:Relationship_best_asymptotic}
\end{align}
where $\mathcal{J}' \triangleq \cup_{n \in \{n \in\mathcal{N}|T_n^*=0\}}\mathcal{J}_n \cup \left(\mathcal{J}_n\backslash\cup_{\{n \in \mathcal{N}|T_n^*=1\}}\mathcal{J}_n\right)$. Here, the near-optimal solution of the service combination probability from Corollary \ref{corollary_1} is also summarized in Algorithm \ref{Al:3}.
\section{Numerical Results} \label{NumericalResults}

In this section, we show numerical results of the \ac{SSP} given by Algorithms \ref{Al:1} and \ref{Al:3} for \ac{EC}-enabled networks. Unless otherwise specified, we consider 10 latency sensitive services\footnote{Note that $10$ service types and $5\sim8$ \acp{BS}'cache sizes (i.e., $N=10$ and $K=5\sim8$) are considered as most existing works on service caching such as \cite{XuCheZho:18,TraChaPom:19,LiZhaJiLi:19,CheHaoHuHosGho:18,WenCuiQueZheJin:20}. However, our algorithm can be applicable for larger $N$ and $K$ with reasonably small execution time such as 0.4982 second with $N=500$ and $K=50$ for Algorithm \ref{Al:3}.} with the same computation workload size, i.e., $N=10$ and $\frac{F_{\text{bs}}}{f_n}=\mu,\, \forall n \in \mathcal{N}$. According to \cite{XuCheZho:18,BiHuaZha:20}, and \cite{RenYuHeLi:19}, we set $K=3$, $\Adensity{bs} = 5\times10^{-4}$ $\text{node/}\text{m}^2$, $\Adensity{u} = 3\times10^{-3}$ $\text{node/}\text{m}^2$, $W = 10\text{MHz}$, $f_{n} = 3\times10^{5}$ $\text{Cycles}$, $\delta=100$, $\omega=1000$, $\Tsize{i} = 420\text{KB}$, $\Tsize{o} = 42\text{KB}$, $p_s = 1$, $\kappa = 30$, $\tau=10^{-4}$, $\alpha=4$, $\Tdead{U} =0.84\text{s}$, $\Tdead{D} =84\text{ms}$, and $\Tdead{Q}=1\text{s}$, unless otherwise specified. We assume that users' service preference follows the Zipf distribution, i.e., $p_n=\frac{n^{\epsilon}}{\sum_{n \in \mathcal{N}} n^{\epsilon}}$, where the Zipf exponent is $\epsilon = 1.1$ \cite{kwak2018hybrid,CuiJia:16}. To assess the performance of the proposed algorithms, we consider the following baseline schemes: 

\begin{figure}[t]
	\begin{center}
		\subfigure[Sucessful uplink transmission, $P_n^{(\text{U})}\left(\textbf{a} \right)$]{
			\psfrag{Delay}[bc][tc][0.85] {Scaling Factor of Target Delays, $R$}
			\psfrag{Probability}[Bc][bc][0.85] {Average SUTP, $\mathbb{E}_{n}\left[P_n^{(\text{U})}\left(\textbf{a} \right) \right]$}
			\psfrag{AlgorithmAlgorithmAlgorithmAlgorithmAlgorithm}[Bl][Bl][0.62] {$\,$Analysis Result (RT and DT)}
			\psfrag{Algorithm}[Bl][Bl][0.62] {$\,$Monte Carlo Simulation (RT and DT)}
			\includegraphics[width=0.8\columnwidth]{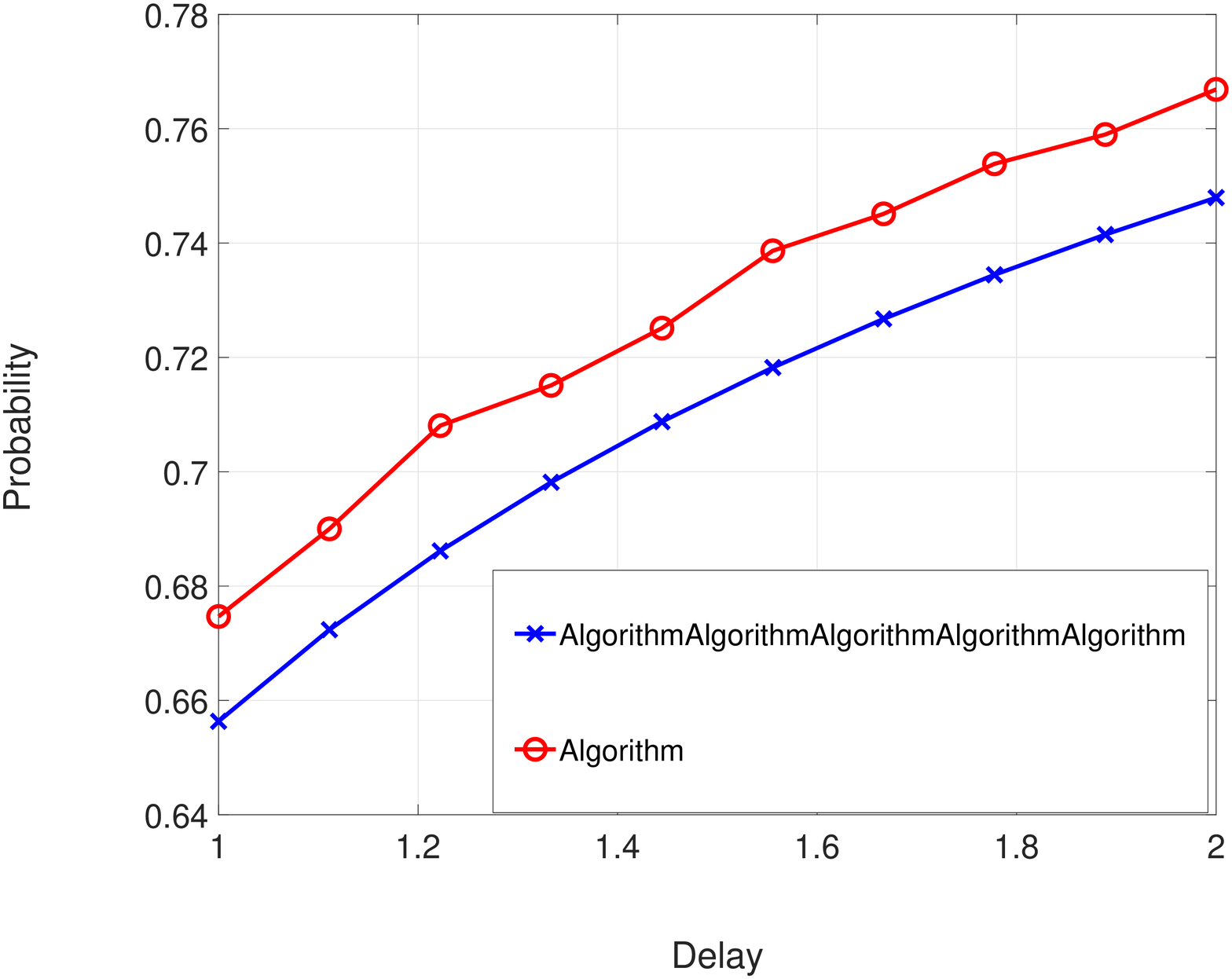}
		}
		\subfigure[Successful downlink transmission, $P_n^{(\text{D})}\left(\textbf{a} \right)$]{
			\psfrag{Delay}[bc][tc][0.8] {Scaling Factor of Target Delays, $R$}
			\psfrag{Probability}[Bc][bc][0.8] {Average SDTP, $\mathbb{E}_{n}\left[P_n^{(\text{D})}\left(\textbf{a} \right) \right]$}
			\psfrag{AlgorithmAlgorithmAlgorithmAlgorithmAlgorithm}[Bl][Bl][0.62] {$\,$Analysis Result (RT and DT)}
			\psfrag{Algorithm}[Bl][Bl][0.62] {$\,$Monte Carlo Simulation (RT and DT)}
			\includegraphics[width=0.8\columnwidth]{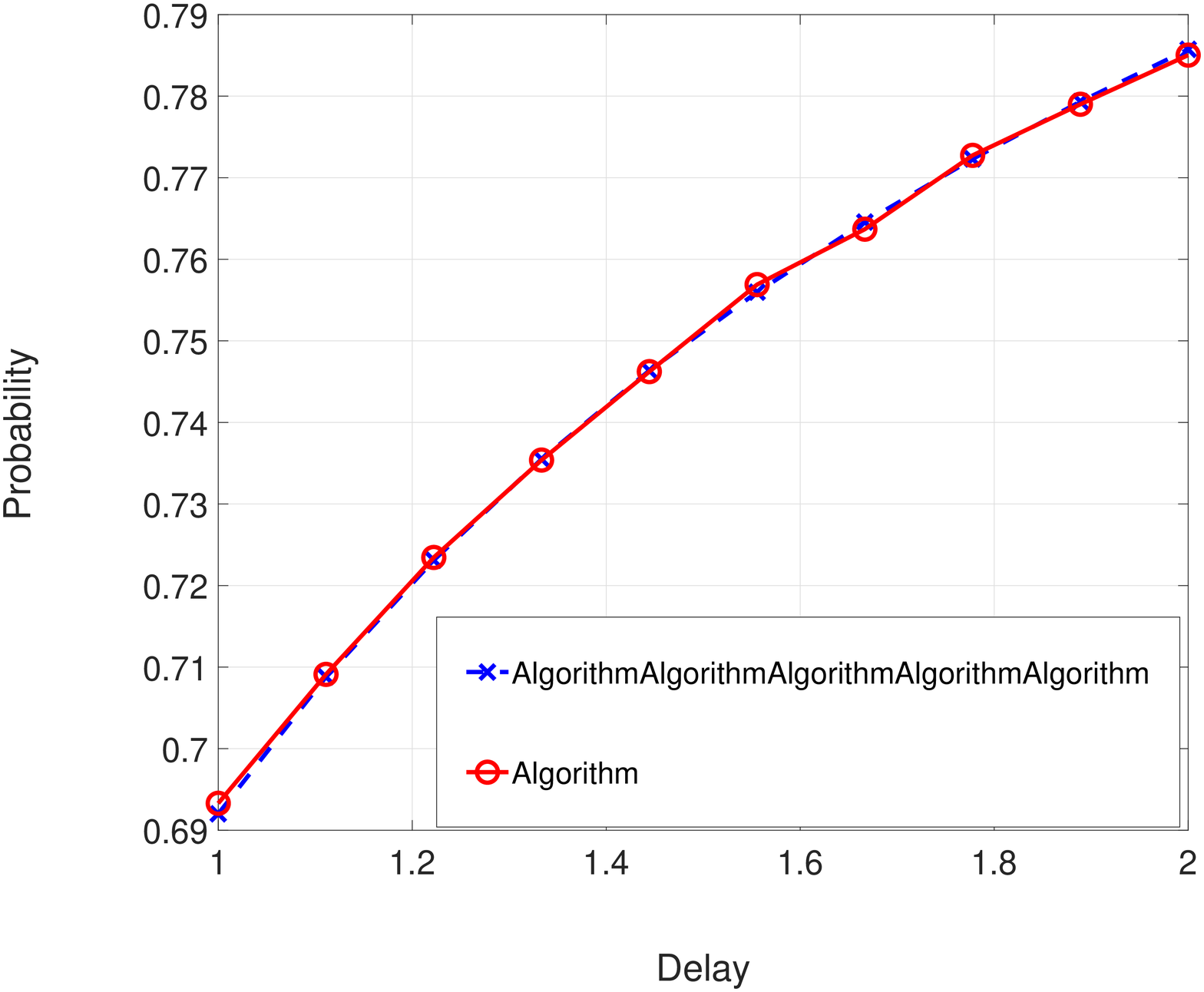}
		}
	\end{center}
	\caption{
		Average \ac{SUTP} $\mathbb{E}_{n}\left[P_n^{(\text{U})}\left(\textbf{a} \right) \right]$ and \ac{SDTP} $\mathbb{E}_{n}\left[P_n^{(\text{D})}\left(\textbf{a} \right) \right]$ as a function of the scaling factor of target delays $R$ with $F_{\text{bs}} = 7.5\times 10^{6}$ $\text{Cycles/s}$.
	}
	\label{fig:Revision_SUTP_SDTP}
\end{figure}

\begin{figure}[t]
	\begin{center}
		\psfrag{Delay}[bc][tc][0.85] {Scaling Factor of Target Delays, $R$}
		\psfrag{Probability}[Bc][bc][0.85] {Average SCPP, $\mathbb{E}_{n}\left[P_{n,k}^{(\text{Q})}\left(\textbf{a},\textbf{b} \right) \right]$}
		\psfrag{AlgorithmAlgorithmAlgorithmAlgorithm}[Bl][Bl][0.62] {$\,$Analysis Result (RT)}
		\psfrag{Algorithm1}[Bl][Bl][0.62] {$\,$Monte Carlo Simulation (RT)}
		\psfrag{Algorithm2}[Bl][Bl][0.62] {$\,$Analysis Result (DT)}
		\psfrag{Algorithm3}[Bl][Bl][0.62] {$\,$Approximate Anaylsis (DT)}
		\psfrag{Algorithm4}[Bl][Bl][0.62] {$\,$Monte Carlo Simulation (DT)}
		\includegraphics[width=0.8\columnwidth]{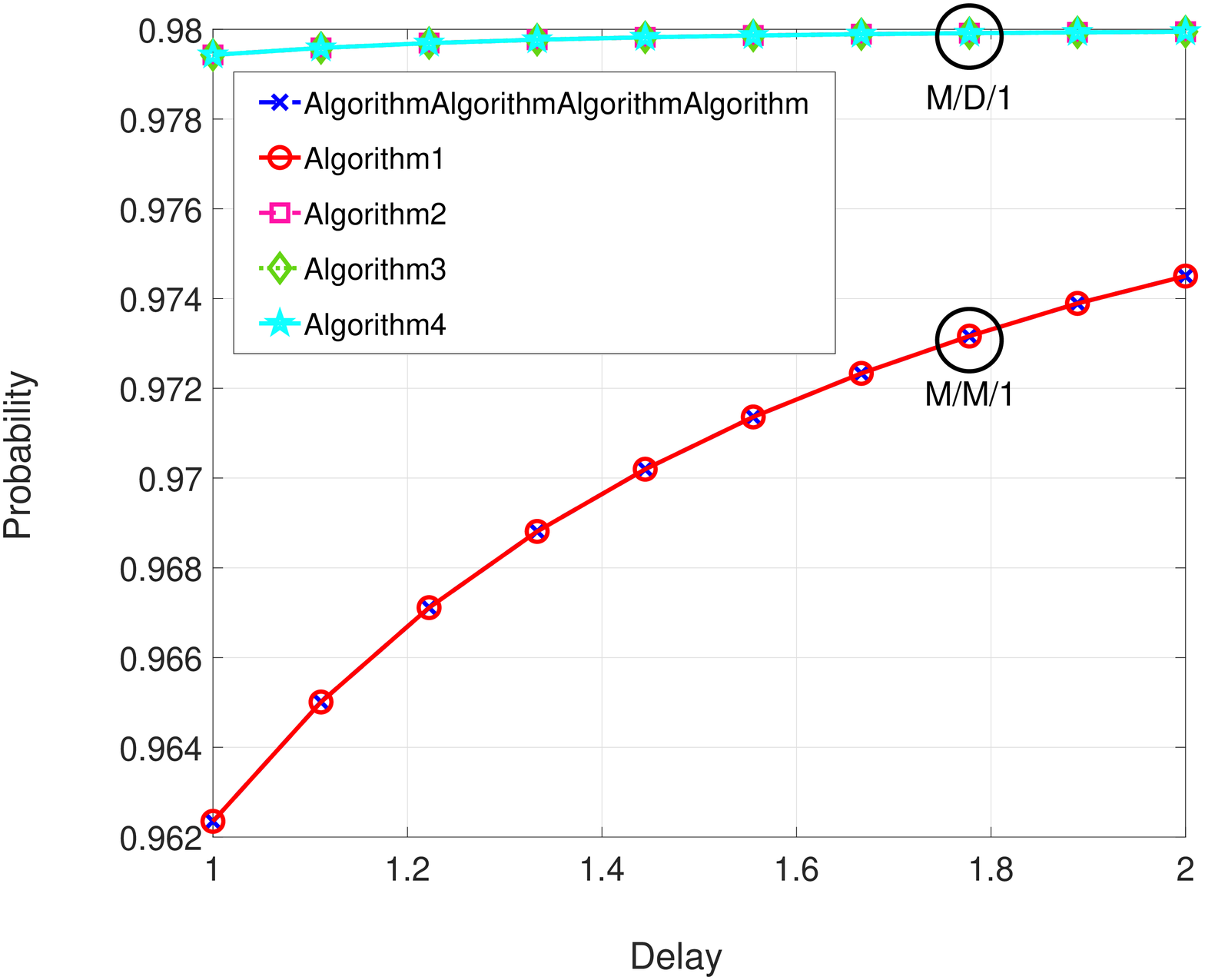}
	\end{center}
	\caption{
		Average \acp{SCPP} $\mathbb{E}_{n}\left[P_{n,\text{m}}^{(\text{Q})}\left(\textbf{a},\textbf{b} \right) \right]$ (RT) and $\mathbb{E}_{n}\left[P_{n,\text{d}}^{(\text{Q})}\left(\textbf{a},\textbf{b} \right) \right]$ (DT) as a function of the scaling factor of target delays $R$ with $F_{\text{bs}} = 7.5\times 10^{6}$ $\text{Cycles/s}$.
	}
	\label{fig:Revision_SCPP}
\end{figure}

\begin{figure}[t!] 
	\begin{center}
		{
			\psfrag{Scaling}[bc][tc][0.8] {Scaling Factor of Target Delays, $R$}
			\psfrag{Probability}[Bc][bc][0.8] {SSP, $\mathcal{P}_{k}\left(\textbf{a},\textbf{b} \right)$}
			\psfrag{AlgorithmAlgorithmAlgorithmAlgorithmAl}[Bl][Bl][0.55] {$\,$Simulation (RT) without Assumption}
			\psfrag{data2}[Bl][Bl][0.55] {$\,$Simulation (RT) with Assumption}
			\psfrag{data3}[Bl][Bl][0.55] {$\,$Anaylsis (RT) with Assumption}
			\psfrag{data4}[Bl][Bl][0.55] {$\,$Simulation (DT) without Assumption}
			\psfrag{data5}[Bl][Bl][0.55] {$\,$Simulation (DT) with Assumption}
			\psfrag{data6}[Bl][Bl][0.55] {$\,$Anaylsis (DT) with Assumption}
			\psfrag{data7}[Bl][Bl][0.55] {$\,$App. Anaylsis (DT) with Assumption}
			\includegraphics[width=0.8\columnwidth]{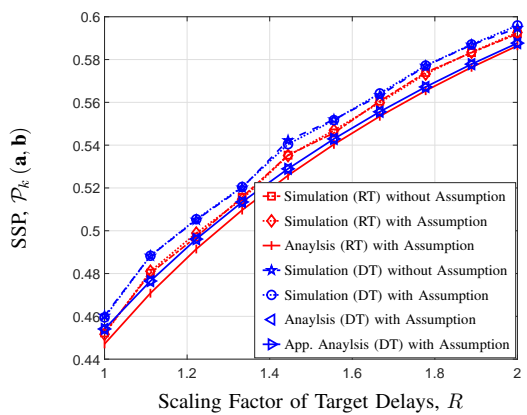}
		}
	\end{center}
	\caption{
		Successful service probabilities, $\tilde{\mathcal{P}}_{\text{m}}$ (RT) and $\tilde{\mathcal{P}}_{\text{d}}$ (DT), as a function of the scaling factor of target delays $R$ with $F_{\text{bs}} = 7.5\times 10^{6}$ $\text{Cycles/s}$.
	}
	\label{fig:approximation_figure}
\end{figure}

\begin{itemize}
	\item \ac{UCPS} scheme : The service caching distribution $\OptVar{a}$ follows the uniform distribution \cite{TamBenNarLat:15}, and the computing resource allocation $\OptVar{b}$ is proportional to users' service preference.
	\item \ac{GCPS} scheme : The service probability $\mathbf{T}$ is determined by users' service preference, i.e., $T_n=p_n$, and $\mathbf{a}$ is determined by the optimal geographic method in \cite{BlaGio:15}. Similar to the \ac{UCPS} scheme, $\OptVar{b}$ is proportional to users' service preference.
	\item \ac{TCPS} scheme : $\OptVar{a}$ is optimized by considering the downlink transmission performance as in \cite{CuiJia:16}, and $\OptVar{b}$ is proportional to users' service preference.
	\item \ac{PCOS} scheme : Each \ac{BS} selects $K$ most popular service software \cite{BasBenDeb:15}. The computation resource allocation $\OptVar{b}$ is optimized using Algorithm \ref{Al:1} for given $\OptVar{a}$.
\end{itemize}
Note that to the best of our knowledge, none of the previous works have jointly optimized $\mathbf{a}$ and $\mathbf{b}$. Therefore, we compare the proposed algorithms with the works above that optimize the content caching probability based on users' service preference. 

Figure \ref{fig:Revision_SUTP_SDTP} shows the impact of target delays on the average \ac{SUTP} in \eqref{eq:SUTP} and the average \ac{SDTP} in \eqref{eq:SDTP}. Here, the scaling factor of the target delays, $R$, is introduced, to investigate the impact of target delays on uplink and downlink transmissions. Specifically, for given initial target delays $\gamma_{\text{o}}^{(\text{U})}$, $\gamma_{\text{o}}^{(\text{Q})}$, and $\gamma_{\text{o}}^{(\text{D})}$, the target delays are determined as $\gamma_n^{(\text{U})} = R\gamma_{\text{o}}^{(\text{U})}$, $\gamma_n^{(\text{Q})} = R\gamma_{\text{o}}^{(\text{Q})}$, and $\gamma_n^{(\text{D})} = R\gamma_{\text{o}}^{(\text{D})}$. From Fig.~\ref{fig:Revision_SUTP_SDTP}(a), we can see that the simulation results of the \ac{SUTP} have the similar trend to the analysis results of the \ac{SUTP}. However, there is a performance gap between the simulation results and the analysis results, caused by the uplink transmission assumption in Section \ref{Analysis}. On the other hand, in the case of \ac{SDTP}, Fig.~\ref{fig:Revision_SUTP_SDTP}(b), we can see that the simulation results show a good agreement with the analysis results as no assumption is used for the analysis.

Figure \ref{fig:Revision_SCPP} shows the average \ac{SCPP} according to $R$. In Fig.~\ref{fig:Revision_SCPP}, `Analysis Result (RT)' refers to the \ac{RT} case in \eqref{eq:SCPP}, `Analysis Result (DT)' refers to the \ac{DT} case in \eqref{eq:SCPP_2}, and `Approximate Analysis (DT)' refers to the approximated SCPP of \ac{DT} case in \eqref{eq:SCPP_2_ap}. From Fig.~\ref{fig:Revision_SCPP}, we can see that there is no performance gap between the approximated analysis result in \eqref{eq:SCPP_2_ap} and the analysis result in \eqref{eq:SCPP_2} for the \ac{DT} case. This is because, as mentioned in Section \ref{System model}, the gap between the original indicator function and the logistic sigmoid function is small. Moreover, from Fig.~\ref{fig:Revision_SCPP}, we can see that the simulation results of \ac{RT} and \ac{DT} cases have the similar trend to the analysis results as no assumption is used for the analysis.

Figure~\ref{fig:approximation_figure} shows the SSPs for RT and DT cases in \eqref{eq:SSP_close} and \eqref{eq:SSP_close_2} according to $R$. Here, the simulation and analysis results `with Assumption' refer to the approximated SSP in \eqref{eq:SSP_ap} while those `without Assumption' refer to the SSP in \eqref{eq:SSP}. From Fig.~\ref{fig:approximation_figure}, we can see that the simulation results of \eqref{eq:SSP} show a good agreement with those of \eqref{eq:SSP_ap}. Therefore, we can see that the performance gap, caused by the independence assumption, is ignorably small. Furthermore, from Fig.~\ref{fig:approximation_figure}, we can see that the simulation results of the \ac{SSP} for both the \ac{RT} and \ac{DT} cases also have a similar trend to the analysis results of the \ac{SSP}.

\begin{figure}[t!]
	\begin{center}
		\subfigure[Random Service Time]{
			\psfrag{Computing}[bc][tc][0.8] {Computing Capability, $F_{\text{bs}}$ [$10^{6}$ Cycles/s]}
			\psfrag{Probability}[Bc][bc][0.8] {SSP, $\mathcal{P}_{\text{m}}\left(\textbf{a},\textbf{b} \right)$}
			\psfrag{AlgorithmAlgorithmAlgorithmAlgo}[Bl][Bl][0.55] {$\,$Proposed (Optimal)}
			\psfrag{data2}[Bl][Bl][0.55] {$\,$Proposed (Near-Optimal)}
			\psfrag{DDDDDD}[Bl][Bl][0.55] {$\,$UCPS}
			\psfrag{data4}[Bl][Bl][0.55] {$\,$GCPS}
			\psfrag{RRRRRR}[Bl][Bl][0.55] {$\,$TCPS}
			\psfrag{data6}[Bl][Bl][0.55] {$\,$PCOS}
			\includegraphics[width=0.8\columnwidth]{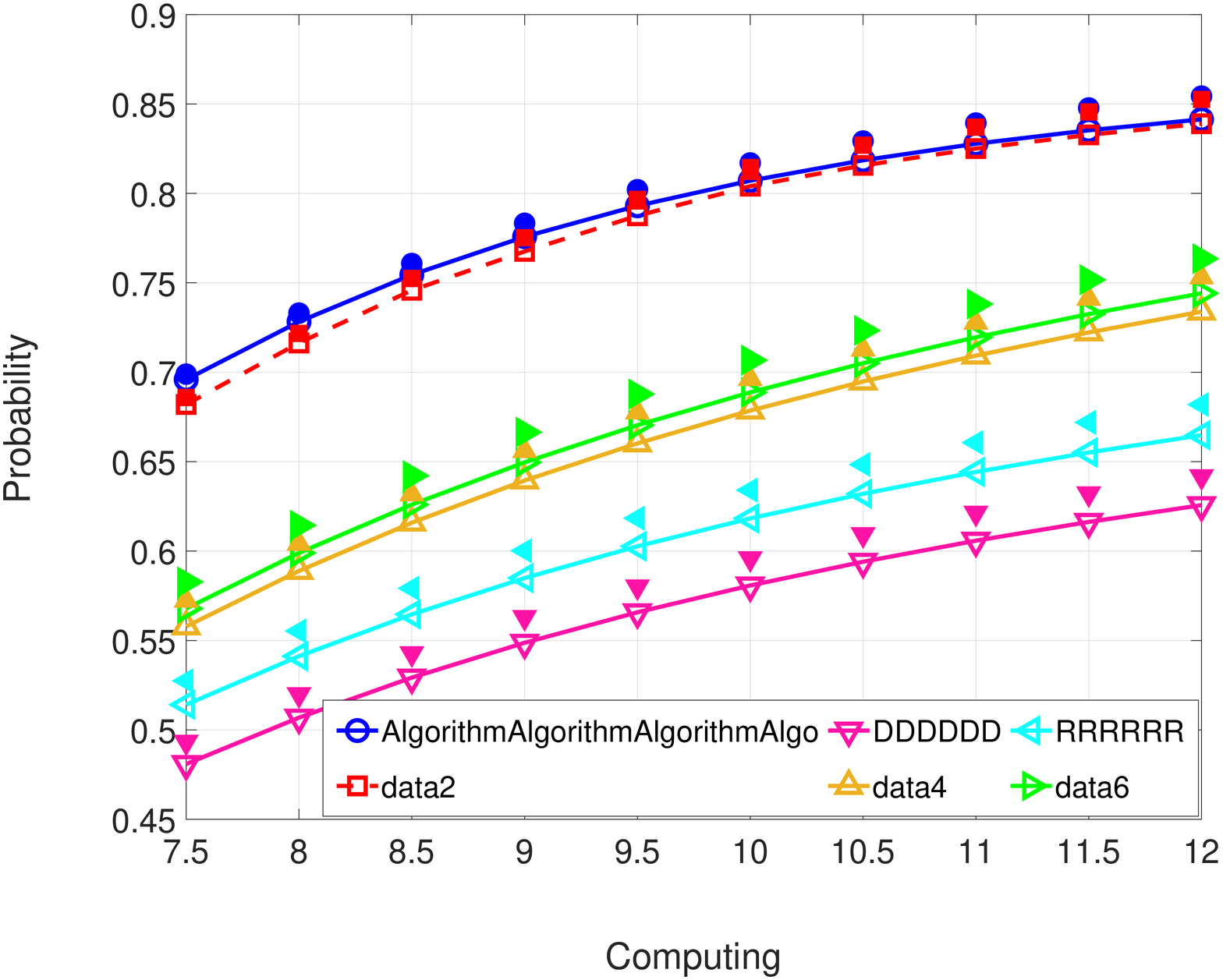}
		}
		\subfigure[Deterministic Service Time]{
			\psfrag{Computing}[bc][tc][0.8] {Computing Capability, $F_{\text{bs}}$ [$10^{6}$ Cycles/s]}
			\psfrag{Probability}[Bc][bc][0.8] {SSP, $\mathcal{P}_{\text{d}}\left(\textbf{a},\textbf{b} \right)$}
			\psfrag{AlgorithmAlgorithmAlgorithmAlgo}[Bl][Bl][0.55] {$\,$Proposed (Optimal)}
			\psfrag{data2}[Bl][Bl][0.55] {$\,$Proposed (Near-Optimal)}
			\psfrag{DDDDDD}[Bl][Bl][0.55] {$\,$UCPS}
			\psfrag{data4}[Bl][Bl][0.55] {$\,$GCPS}
			\psfrag{RRRRRR}[Bl][Bl][0.55] {$\,$TCPS}
			\psfrag{data6}[Bl][Bl][0.55] {$\,$PCOS}
			\includegraphics[width=0.8\columnwidth]{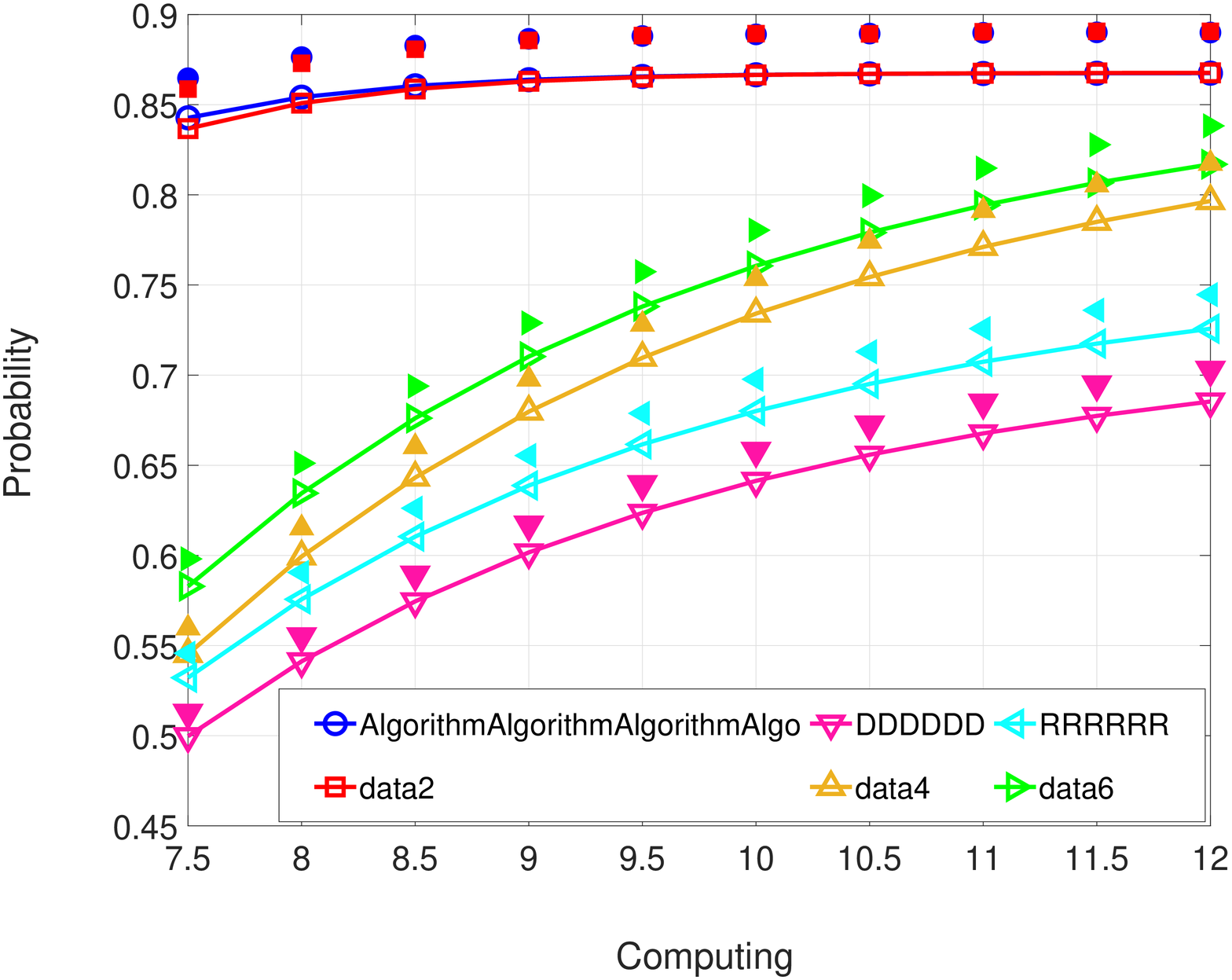}
		}
	\end{center}
	\caption{	
		Successful service probabilities, $\tilde{\mathcal{P}}_{\text{m}}$ (RT) and $\tilde{\mathcal{P}}_{\text{d}}$ (DT), as a function of the computing capability $F_{\text{bs}}$ with $K=8$. The filled symbols represent the simulation results of the \ac{SSP} for two cases.
	}
	\label{fig:computing}
\end{figure}

\begin{figure}[t!]
	\begin{center}
		\subfigure[Random Service Time]{
			\psfrag{Size}[bc][tc][0.8] {BSs' Service Cache Size, $K$}
			\psfrag{Probability}[Bc][bc][0.8] {SSP, $\mathcal{P}_{\text{m}}\left(\textbf{a},\textbf{b} \right)$}
			\psfrag{AlgorithmAlgorithmAlgorithmAlgorithm}[Bl][Bl][0.55] {$\,$Proposed Solution}
			\psfrag{data2}[Bl][Bl][0.55] {$\,$UCPS}
			\psfrag{data3}[Bl][Bl][0.55] {$\,$GCPS}
			\psfrag{data4}[Bl][Bl][0.55] {$\,$TCPS}
			\psfrag{data5}[Bl][Bl][0.55] {$\,$PCOS}
			\includegraphics[width=0.8\columnwidth]{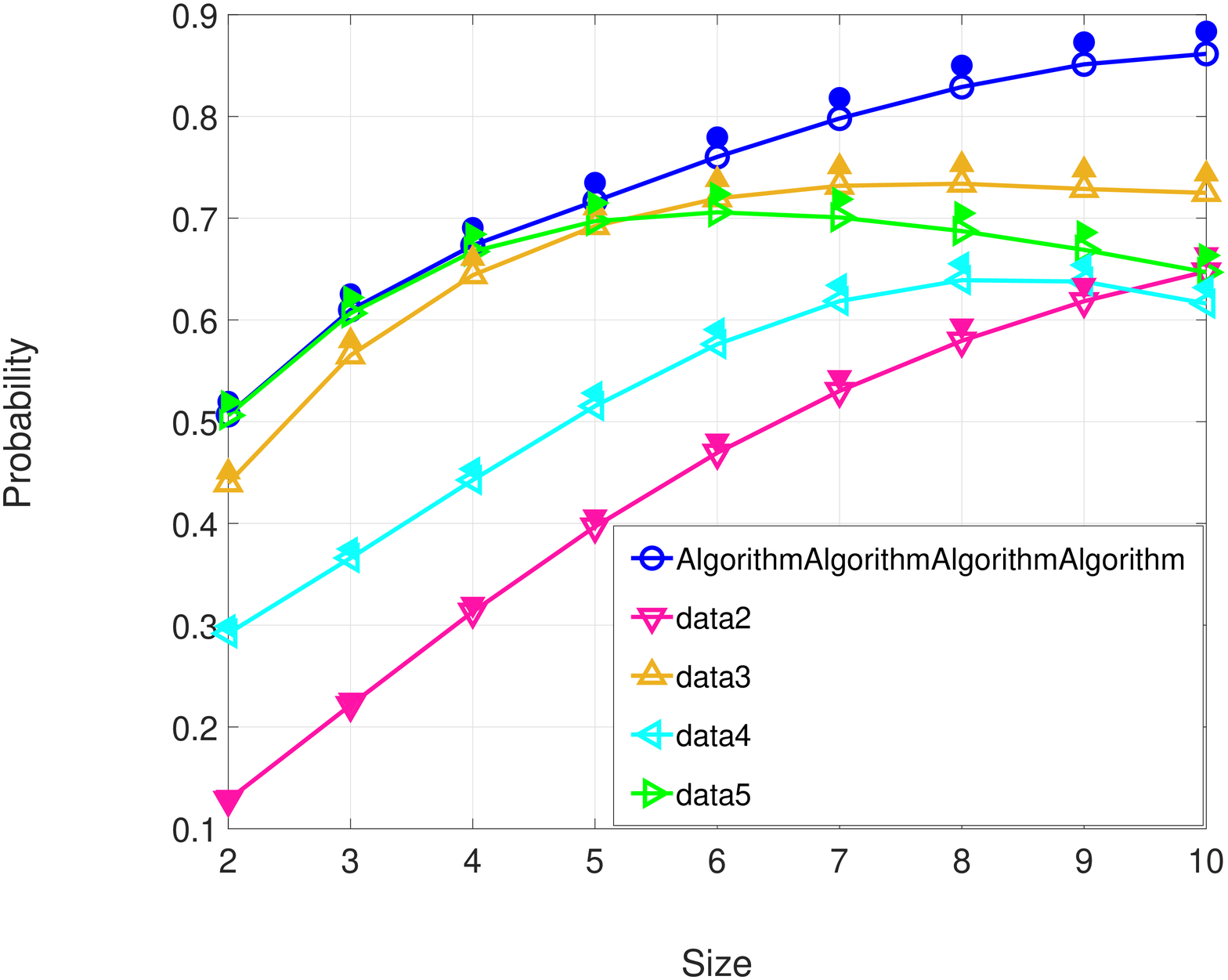}
		}
		\subfigure[Deterministic Service Time]{
			\psfrag{Size}[bc][tc][0.8] {BSs' Service Cache Size, $K$}
			\psfrag{Probability}[Bc][bc][0.8] {SSP, $\mathcal{P}_{\text{d}}\left(\textbf{a},\textbf{b} \right)$}
			\psfrag{AlgorithmAlgorithmAlgorithmAlgorithm}[Bl][Bl][0.55] {$\,$Proposed Solution}
			\psfrag{data2}[Bl][Bl][0.55] {$\,$UCPS}
			\psfrag{data3}[Bl][Bl][0.55] {$\,$GCPS}
			\psfrag{data4}[Bl][Bl][0.55] {$\,$TCPS}
			\psfrag{data5}[Bl][Bl][0.55] {$\,$PCOS}
			\includegraphics[width=0.8\columnwidth]{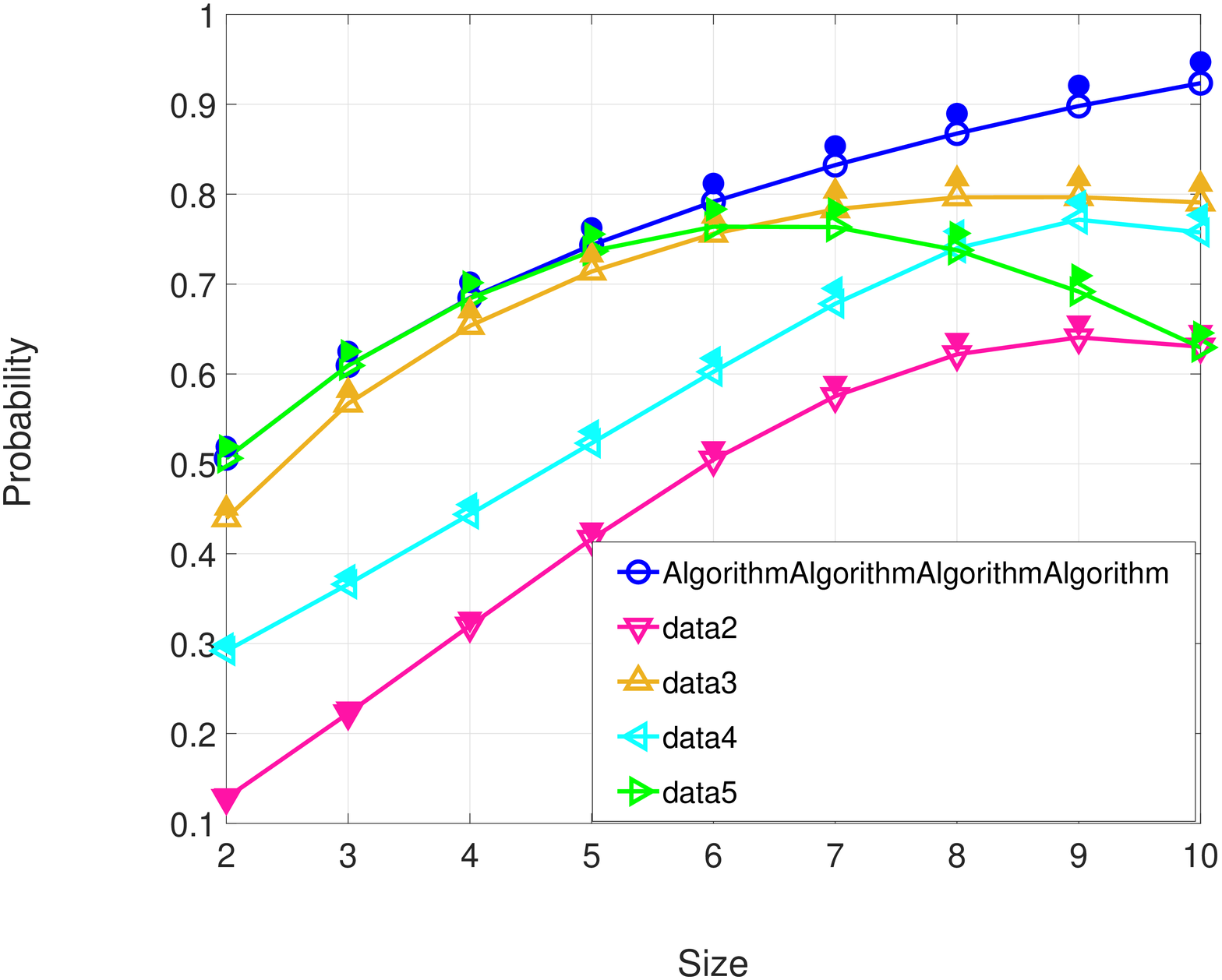}
		}
	\end{center}
	\caption{
		Successful service probabilities, $\tilde{\mathcal{P}}_{\text{m}}$ (RT) and $\tilde{\mathcal{P}}_{\text{d}}$ (DT), as a function of the \acp{BS}' service cache size $K$ with $F_{\text{bs}} = 1.2\times10^{7}$ $\text{Cycles/s}$. The filled symbols represent the simulation results of the \ac{SSP} for two cases.
	}
	\label{fig:cachingsize}
\end{figure}

Figure \ref{fig:computing} shows the \ac{SSP} for \ac{EC}-enabled networks versus the computing capability of the \ac{EC} server, $F_{\text{bs}}$. Here, the optimal and near-optimal solutions are obtained by Algorithms \ref{Al:1} and \ref{Al:3}, respectively. We can see that our proposed solutions outperform the baseline schemes. Moreover, we can see that the performance gap between the optimal solution and near-optimal solution decreases with $F_{\text{bs}}$. In other words, the near-optimal solution achieves reliable performance especially when $F_{\text{bs}}$ is high. In addition, we can see that the \ac{SSP} of each scheme converges to different points with increasing $F_{\text{bs}}$. This is because the \ac{SCPP} is always $1$ due to the sufficiently high computing capability of the \ac{EC} server $F_{\text{bs}}$, while the \ac{SUTP} and \ac{SDTP} have a different values depending on the service caching distribution $\OptVar{a}$.  

\begin{figure}[t!]
	\begin{center}
		\subfigure[SSP when $K=30$]{
			\psfrag{Computing}[bc][tc][0.8] {Computing Capability, $F_{\text{bs}}$ [$10^{7}$ Cycles/s]}
			\psfrag{Probability}[Bc][bc][0.8] {SSP, $\mathcal{P}_{k}\left(\textbf{a},\textbf{b} \right)$}
			\psfrag{GeneralGeneralGen}[Bl][Bl][0.55] {$\,$Proposed (DT)}
			\psfrag{R2}[Bl][Bl][0.55] {$\,$Proposed (RT)}
			\psfrag{DDDDDDDD}[Bl][Bl][0.55] {$\,$GCPS(DT)}
			\psfrag{R4}[Bl][Bl][0.55] {$\,$GCPS(RT)}
			\psfrag{RRRRRRRR}[Bl][Bl][0.55] {$\,$TCPS(DT)}
			\psfrag{R6}[Bl][Bl][0.55] {$\,$TCPS(RT)}
			\psfrag{EEEEEEEE}[Bl][Bl][0.55] {$\,$PCOS(DT)}
			\psfrag{R8}[Bl][Bl][0.55] {$\,$PCOS(RT)}
			\includegraphics[width=0.8\columnwidth]{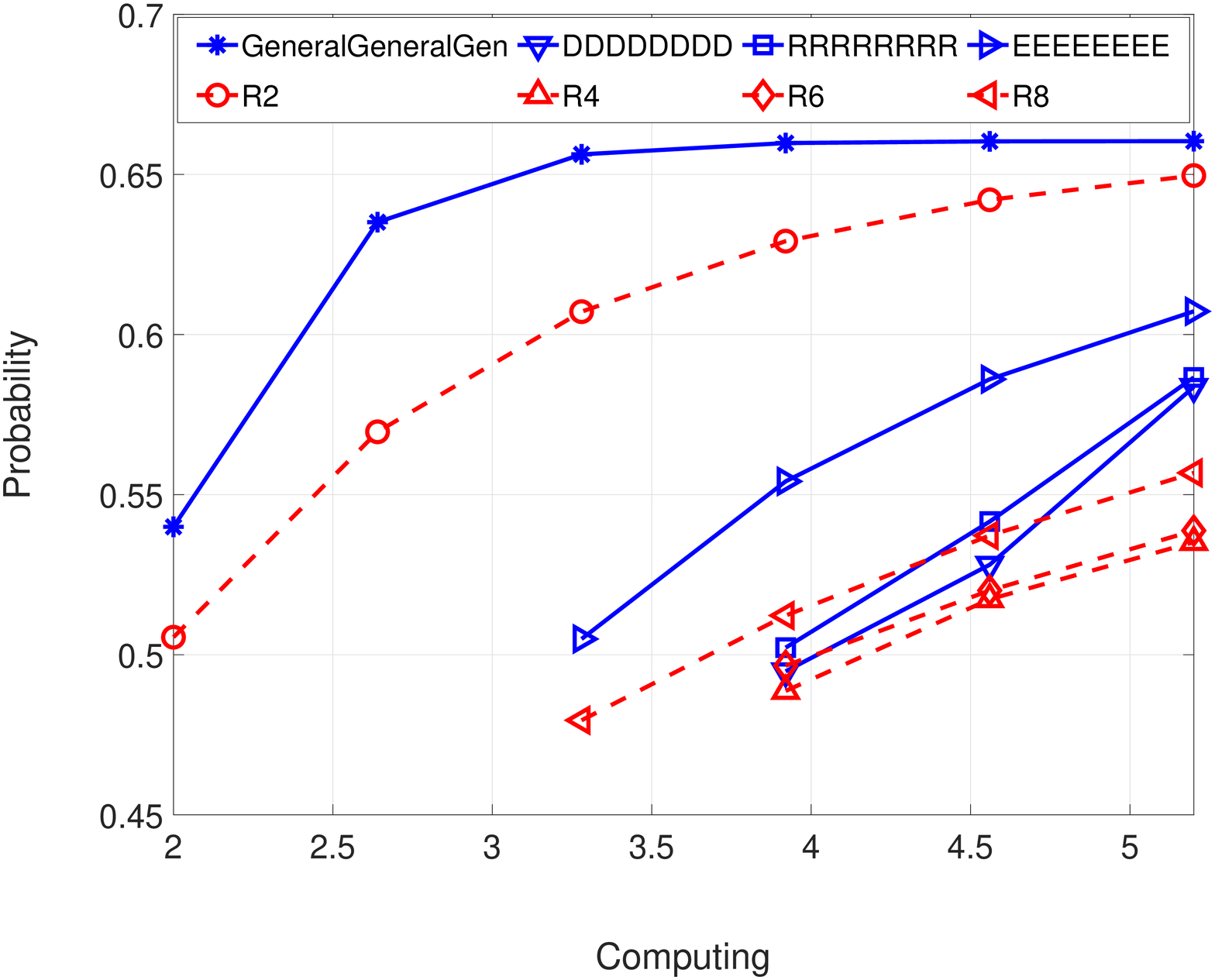}
		}
		\subfigure[SSP when $F_{\text{bs}}=4\times10^{7}$ $\text{Cycles/s}$]{
			\psfrag{Cache}[bc][tc][0.8] {BSs' Service Cache Size, $K$}
			\psfrag{Probability}[Bc][bc][0.8] {SSP, $\mathcal{P}_{k}\left(\textbf{a},\textbf{b} \right)$}
			\psfrag{GeneralGeneralGeneral}[Bl][Bl][0.55] {$\,$Proposed (DT)}
			\psfrag{R2}[Bl][Bl][0.55] {$\,$Proposed (RT)}
			\psfrag{DDDDDDDD}[Bl][Bl][0.55] {$\,$UCPS(DT)}
			\psfrag{R4}[Bl][Bl][0.55] {$\,$UCPS(RT)}
			\psfrag{RRRRRRRR}[Bl][Bl][0.55] {$\,$TCPS(DT)}
			\psfrag{R6}[Bl][Bl][0.55] {$\,$TCPS(RT)}
			\psfrag{EEEEEEEE}[Bl][Bl][0.55] {$\,$GCPS(DT)}
			\psfrag{R8}[Bl][Bl][0.55] {$\,$GCPS(RT)}
			\psfrag{FFFFFFFF}[Bl][Bl][0.55] {$\,$PCOS(DT)}
			\psfrag{R10}[Bl][Bl][0.55] {$\,$PCOS(RT)}
			\includegraphics[width=0.8\columnwidth]{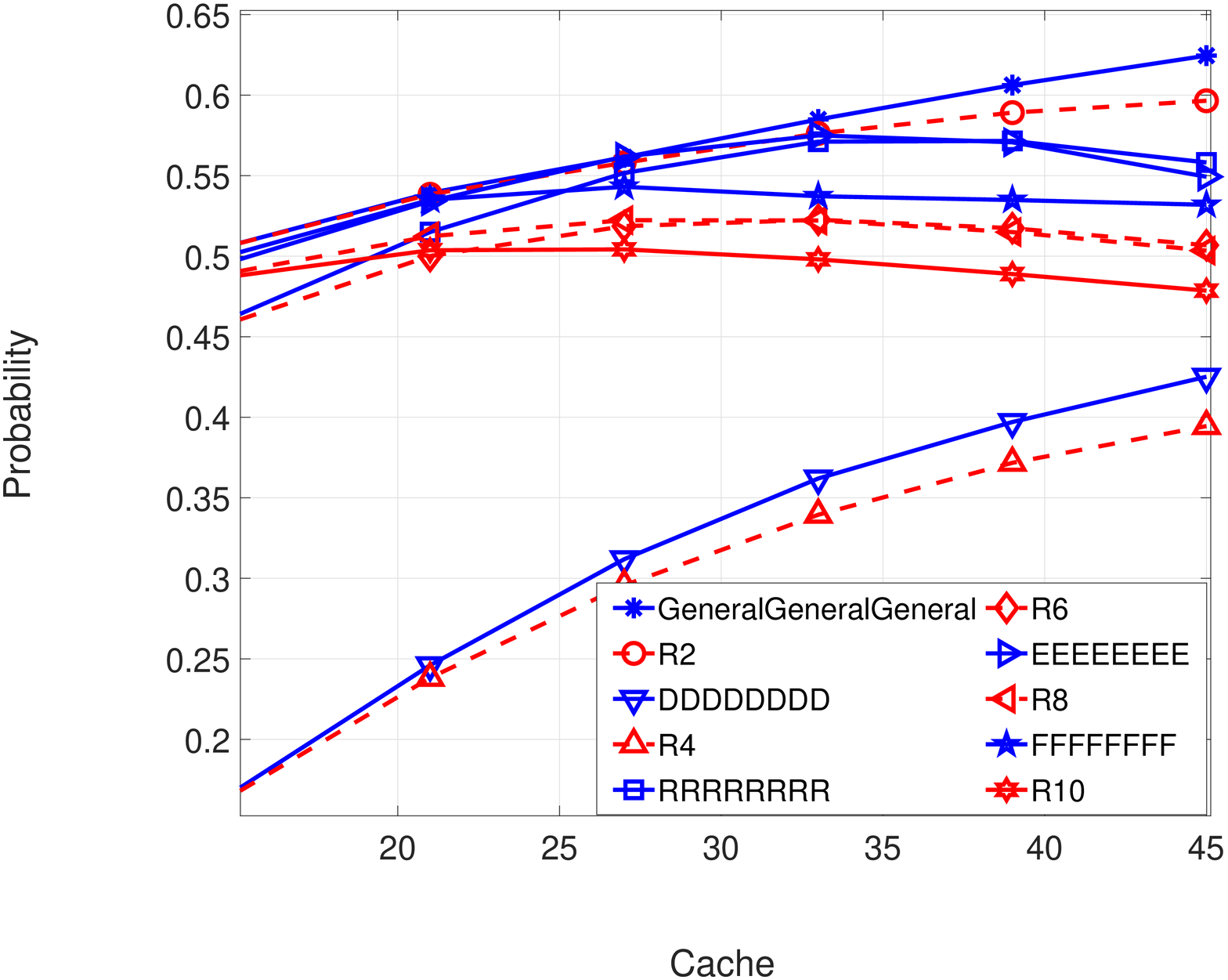}
		}
	\end{center}
	\caption{
		Successful service probabilities, $\tilde{\mathcal{P}}_{\text{m}}$ (RT) and $\tilde{\mathcal{P}}_{\text{d}}$ (DT), as a function of the \acp{BS}' service cache size $K$ and the computing capability $F_{\text{bs}}$ with $N=100$.
	}
	\label{fig:dddddd}
\end{figure}

From Fig. \ref{fig:approximation_figure}. and Fig. \ref{fig:computing}., we can see that the \ac{SSP} increases with $R$ and $F_{\text{bs}}$. Moreover, the \ac{SSP} in the \ac{DT} case is always higher than the \ac{SSP} in the \ac{RT} case due to the randomness of computation time in the \ac{RT} case. 

Figure \ref{fig:cachingsize} shows the \ac{SSP} for \ac{EC}-enabled networks versus the \acp{BS}' service caching storage $K$. From Fig.~\ref{fig:cachingsize}, we can see that the SSP of each baseline scheme first increases and then decreases with the BSs' cache size $K$. For the small $K$, as $K$ increases, there can be larger number of \acp{BS}, which store the requested service software, and this leads to the high SUTP and SDTP. Consequently, the SSP increases. However, as $K$ keeps increasing, the computing capability assigned for each service software at a EC server decreases due to the limited computing capability. Since baseline schemes do not jointly consider the service caching distribution and the computing resource allocation, the computation time monotonically increases with $K$, which results in low SCPP compared to the SUTP and the SDTP. As a result, the SSP decreases. On the other hand, from Fig.~\ref{fig:cachingsize}, we can see that the SSP of the proposed solution keeps increasing with $K$ due to the joint optimization of service caching distribution and the computing resource allocation.

Furthermore, from Fig.~\ref{fig:computing} and Fig.~\ref{fig:cachingsize}, we can see that the proposed solutions outperform the baseline schemes even when the SSP is high, e.g., when $F_{\text{bs}}\geq10^{7}\,\text{Cycles/s}$ and $K\geq9$.

Figure \ref{fig:dddddd} shows the impact of computing capability of the \ac{EC} server $F_{\text{bs}}$ and the \acp{BS}' service caching storage $K$ on the \ac{SSP} with large number of services $(N=100)$. Since it is hard to obtain the optimal solution due to the computational complexity, we only show the near-optimal solution obtained by Algorithm \ref{Al:3}. From Fig. \ref{fig:dddddd}, we can observe that the proposed solution outperforms all the baseline schemes. Specifically, in Fig. \ref{fig:dddddd}, when $F_{\text{bs}}$ is low, some baseline schemes are not feasible as they cannot satisfy the stability condition of the service queue. However, the proposed solution satisfies the stability condition of the service queue regardless of $F_{\text{bs}}$. Moreover, as the \ac{SSP} increases with $F_{\text{bs}}$ and $K$, we can see that the near-optimal solution can be applied to \ac{EC}-enabled networks with large number of services.

\section{Conclusion}\label{Conclusion}
In this paper, we develop an efficient service caching and a computing resource allocation in \ac{EC} enabled networks with multiple types of latency sensitive services. Specifically, we derive the closed form expression of the \ac{SSP} in both the \ac{RT} and \ac{DT} cases. Then, to maximize the \ac{SSP}, we formulate optimization problems of the service caching distribution and the computing resource allocation in both cases, which are challenging nonconvex problems. We propose an iterative algorithm to obtain a stationary point of the \ac{SSP} maximization problem. Then, in the high computing capability region, we formulate the asymptotic \ac{SSP} maximization problem and obtain the stationary point of the asymptotic \ac{SSP} maximization problem. Furthermore, we develop an iterative algorithm with low complexity to obtain a near-optimal solution of the \ac{SSP} maximization problem. Finally, from numerical simulations, we show that the proposed solutions in both the general and high computing capability regions achieve higher \ac{SSP} than that of the baseline schemes. Moreover, the near-optimal solution in the high computing capability region achieves a reliable performance compared to the optimal solution in the general region. Then, we show that the \ac{SSP} increases with the target delays, the computing capability of \ac{EC} servers, and \acp{BS}' service cache size in both the \ac{RT} and \ac{DT} cases. We also show the near-optimal solution can be applied for the \ac{EC} enabled networks that have large number of services.
\begin{appendix}
	\subsection{Proof of Lemma~\ref{lem:Op_b}}
	By relaxing computing constraints in \eqref{eq:C_4}, we obtain Lagrange function $L(\OptVar{b}_j,\eta_j),\,j \in \mathcal{J}$ by
	\begin{align}
	L(\OptVar{b}_j,\eta_j) =& \sum_{n \in \mathcal{N}_j}\frac{p_nT_{n,\text{m}}^{(r)}a_{j,\text{m}}^{(r)}e^{\lambda_{n,\text{m}}^{(r)}\Tdead{Q}-b_{n,j}J_n}  }{D_n\left(T_{n,\text{m}}^{(r)}\right)^2+E_nT_{n,\text{m}}^{r}+A_n^2} \nonumber \\
	&-\frac{1}{\omega}\phi_{n,j}\left(\mathbf{a}_{\text{m}}^{(r)},b_{n,j}\right)+\eta_j\left(\sum_{j\in \mathcal{J}_n} b_{n,j}-1  \right),
	\end{align}
	where $\eta_j$ is the Lagrange multiplier \ac{w.r.t.} the constraint in \eqref{eq:C_4}. Then, we obtain the derivative of the Lagrange function given by
	\begin{align}
	\frac{\partial L(\OptVar{b}_j,\eta_j)}{\partial b_{n,j}} = &- \frac{J_np_nT_{n,\text{m}}^{(r)}a_{j,\text{m}}^{(r)}e^{\lambda_{n,\text{m}}^{(r)}\Tdead{Q}-b_{n,j}J_n}  }{D_n\left(T_{n,\text{m}}^{r}\right)^2+E_nT_{n,\text{m}}^{r}+A_n^2} \nonumber \\
	&-\frac{\mu_n}{\omega}e^{\phi_{n,j}\left(\mathbf{a}_{\text{m}}^{(r)},b_{n,j}\right)} +\eta_j, j \in \mathcal{J}.
	\label{eq:Diff_Lagrange}
	\end{align}
	We obtain the \ac{KKT} conditions by \eqref{eq:C_3}, \eqref{eq:C_4}, and
	\begin{align}
	\frac{\partial L(\mathbf{b}_j,\eta_j)}{\partial b_{n,j}}=0, \quad j \in \mathcal{J}.
	\label{eq:KKT condition}
	\end{align} 
	Since the strong condition holds, by substituting \eqref{eq:Diff_Lagrange} into \eqref{eq:KKT condition}, we obtain 
	\begin{align}
	G_{n,j}(b_{n,j}) =  \eta_j,
	\end{align}
	where $G_{n,j}(b_{n,j})$ is given by
	\begin{align}
	G_{n,j}(b_{n,j}) =&  \frac{J_np_nT_{n,\text{m}}^{(r)}a_{j,\text{m}}^{(r)}e^{\lambda_{n,\text{m}}^{(r)}\Tdead{Q}-b_{n,j}J_n}  }{D_n\left(T_{n,\text{m}}^{r}\right)^2+E_nT_{n,\text{m}}^{r}+A_n^2}  \nonumber \\
	&+\frac{\mu_n}{\omega}e^{\phi_{n,j}\left(\mathbf{a}_{\text{m}}^{(r)},b_{n,j}\right)}.
	\label{eq:KKT_result}
	\end{align}
	By considering \eqref{eq:C_3}, \eqref{eq:C_4}, and \eqref{eq:KKT_result}, we represent the optimal solution $\bar{b}_{n,j,\text{m}}^{(r+1)}$, by \eqref{eq:op_server_splitting}.
	\label{app:Op_b}
	\subsection{Proof of Lemma~\ref{lemma: op_sub_problem}}
	First, by relaxing the constraint in \eqref{eq:C_4}, we obtain the Lagrange function $L(\OptVar{b}_j,\zeta_j),j \in \mathcal{J}$ by
	\begin{align}
	L(\OptVar{b}_{j},\zeta_j) = \sum_{n \in \mathcal{N}_j}&\frac{p_nT_n^{*}e^{\lambda_{n}^{*}\Tdead{Q}-b_{n,j}J_n}}{D_n\left(T_n^{*}\right)^2+E_nT_n^{*}+A_n^2} \nonumber \\
	&+\zeta_j\left(1-\sum_{j\in \mathcal{J}_n} b_{n,j}  \right)\hspace{-1mm},
	\end{align}
	where $\zeta_j$ is the Lagrange multiplier \ac{w.r.t.} the constraint in \eqref{eq:C_4}. Then, we obtain the derivative of  Lagrange function given by
	\begin{align}
	&\frac{\partial L(\OptVar{b}_{j},\zeta_j)}{\partial b_{n,j}} = \frac{-J_np_nT_n^{*} e^{\lambda_{n}^{*}\Tdead{Q}-b_{n,j}J_n}}{D_n\left(T_n^{*}\right)^2+E_nT_n^{*}+A_n^2} +\zeta_j, \, j \in \mathcal{J}.
	\label{eq:Diff_Lagrange_special}
	\end{align}
	We obtain the KKT conditions by \eqref{eq:C_3}, \eqref{eq:C_4}, and
	\begin{align}
	\frac{\partial L(\mathbf{b}_j,\zeta_j)}{\partial b_{n,j}}=0, \quad j \in \mathcal{J}.
	\label{eq:KKT condition_special}
	\end{align} 
	Since the strong condition holds, by substituting \eqref{eq:Diff_Lagrange_special} into \eqref{eq:KKT condition_special}, we obtain $b_{n,j},\,j \in \mathcal{J}$ by
	\begin{align}
	\bar{b}_{n,j,\text{m}}= \frac{1}{J_n}\left\{\lambda_{n}^{*}\Tdead{Q}+\log\left(\frac{\frac{J_np_nT_n^*}{D_n(T_n^*)^2+E_nT_n^*+A_n^2}}{\hat{\zeta_j}} \right)\right\}\hspace{-1mm}.
	\end{align}
	By considering \eqref{eq:C_3}, \eqref{eq:C_4}, and \eqref{eq:KKT condition_special},  for $n \in \mathcal{N}_j,j \in \mathcal{J}$, we represent the optimal solution $\bar{b}_{n,j,\text{m}}$, which is given by
	\begin{align}
	\bar{b}_{n,j,\text{m}} = \max&\left\{\left(\log\left(\frac{J_np_nT_n^*}{\hat{\zeta_j}\left(D_n(T_n^*)^2+E_nT_n^*+A_n^2\right)} \right)\right.\right. \nonumber \\
	&\quad\left.\left. \times\frac{1}{J_n}+\frac{\lambda_{n}^{*}}{\mu_n}\right), \frac{\lambda_{n}^{*}}{\mu_n} \right \}.
	\label{eq:Op_b_asymptotic}
	\end{align}
	Next, we suppose that for all $n \in \mathcal{N}_j$ and $j \in \mathcal{J}$, the optimal solution $\bar{b}_{n,j,\text{m}}$ is constructed only by the first term in the max function of \eqref{eq:Op_b_asymptotic}, which is represented by
	\begin{align}
	\bar{b}_{n,j,\text{m}}
	=
	\frac{1}{J_n}\log\left(\frac{J_np_n(T_n^*)^2}{\hat{\zeta_j}\left(D_nT_n^*+E_nT_n^*+A_n^2\right)} \right)
	+
	\frac{\lambda_{n}^{*}}{\mu_n}.
	\label{eq:Op_b_asymptotic_suppose}
	\end{align}
	Then, we can obtain the closed form expression of the Lagrangian multiplier $\hat{\zeta_j}^{*}$, which satisfies $\sum_{n \in \mathcal{N}_j} \bar{b}_{n,j,\text{m}}  = 1, \,j \in \mathcal{J}$, by
	\begin{align}
	\hat{\zeta_j}^{*} = e^{\frac{-1+\sum_{n \in \mathcal{N}_j}\frac{1}{J_n} \left(\lambda_{n}^{*}\Tdead{Q}+\log\left(\frac{J_np_nT_n^*}{D_n(T_n^*)^2+E_nT_n^*+A_n^2} \right) \right)}{\sum_{n \in \mathcal{N}_j}\frac{1}{J_n}}}. 
	\label{eq:Lagrangian_multiplier_asymptotic}
	\end{align}
	By substituting $\hat{\zeta_j}^{*}$ into $\hat{\zeta_j}$ in \eqref{eq:Op_b_asymptotic}, we can rewrite the optimal solution $b_{n,j}^*$ as follows.
	\begin{align}
	\bar{b}_{n,j,\text{m}} = \max&\left\{\left(\log\left(\frac{J_np_nT_n^*}{\hat{\zeta_j}^*\left(D_n(T_n^*)^2+E_nT_n^*+A_n^2\right)} \right)\right.\right. \nonumber \\
	&\left.\left.\quad\,\, \times \frac{1}{J_n}+\frac{\lambda_{n}^{*}}{\mu_n}\right), \frac{\lambda_{n}^{*}}{\mu_n} \right \}, n \in \mathcal{N}_j,j \in \mathcal{J}.
	\label{eq:Op_b_asymptotic_2}
	\end{align}
	Then, we can rewrite $\log\left(\frac{J_np_nT_n^*}{\hat{\zeta_j}^*\left(D_n(T_n^*)^2+E_nT_n^*+A_n^2\right)} \right),\, n \in \mathcal{N}$ as follow
	\begin{align}
	&\log\left(\frac{J_np_nT_n^*}{\hat{\zeta_j}^*\left(D_n(T_n^*)^2+E_nT_n^*+A_n^2\right)} \right) \nonumber \\
	&\quad=\frac{1-\sum_{n \in \mathcal{N}_j}\frac{\lambda_{n}^{*}}{\mu_n}+\sum_{k \in \mathcal{N}_j}\frac{G_{n,k}}{J_k}}{\sum_{n \in \mathcal{N}_j}\frac{1}{J_n}}, 
	\label{eq:con_Op_b_asymptotic}
	\end{align}
	where $G_{n,k} = \log\left(\frac{J_np_nT_n^*}{\left(D_n(T_n^*)^2+E_nT_n^*+A_n^2\right)} \right)-\log\left(\frac{J_kp_kT_k^*}{\left(D_k(T_k^*)^2+E_kT_k^*+A_k^2\right)} \right)$.
	
	Since $\frac{\log\left(\frac{J_np_nT_n^*}{\hat{\zeta_j}^*\left(D_n(T_n^*)^2+E_nT_n^*+A_n^2\right)} \right)}{J_n} \leq 1-\sum_{n \in \mathcal{N}_j} \frac{\lambda_{n}^{*}}{\mu_n} +\frac{1}{J_n}\log\left(\zeta_j\right)-\frac{\lambda_{n}^{*}}{\mu_n}$, and $\sum_{k \in \mathcal{N}_j}\frac{G_{n,k}}{J_k}\geq -1+\sum_{n \in \mathcal{N}_j} \frac{\lambda_{n}^{*}}{\mu_n}$, the condition in \eqref{eq:con_Op_b_asymptotic} is always greater than $0$. Therefore, the optimal solution in \eqref{eq:Op_b_asymptotic_2} is always greater than $\frac{\lambda_{n}^{*}}{\mu_n}$. Therefore, we can remove the max function and obtain the closed form optimal solution given by \eqref{eq:op_special_b}.
	\label{app:op_sub_problem}
\end{appendix}


\begin{thebibliography}{99}
\bibitem{KimChoCuiLee:20}
M.~Kim, H.~Cho, Y.~Cui, and J.~Lee, ``Service caching and computation resource
allocation for large-scale edge computing-enabled networks,'' in \emph{Proc.
	IEEE Glob. Commun. Conf. (GLOBECOM)}, Taipei, Taiwan, Dec. 2020, pp. 1--6.

\bibitem{MaoYouZha:17}
Y.~Mao, C.~You, J.~Zhang, K.~Huang, and K.~B. Letaief, ``A survey on mobile
edge computing: The communication perspective,'' \emph{{IEEE} Commun. Surveys
	Tuts.}, vol.~19, no.~4, pp. 2322--2358, Aug. 2017.

\bibitem{XuCheZho:18}
J.~Xu, L.~Chen, and P.~Zhou, ``Joint service caching and task offloading for
mobile edge computing in dense networks,'' in \emph{Proc. IEEE Conf. Comput.
	Commun. (INFOCOM)}, Honolulu, HI, USA, Apr. 2018, pp. 1--9.

\bibitem{TraPom:18}
T.~X. Tran and D.~Pompili, ``Joint task offloading and resource allocation for
multi-server mobile-edge computing networks,'' \emph{{IEEE} Trans. Veh.
	Technol.}, vol.~68, no.~1, pp. 856--868, Nov. 2018.

\bibitem{LiuBenDebPoo:19}
C.-F. Liu, M.~Bennis, M.~Debbah, and H.~V. Poor, ``Dynamic task offloading and
resource allocation for ultra-reliable low-latency edge computing,''
\emph{IEEE Trans. Commun.}, vol.~67, no.~6, pp. 4132--4150, Feb. 2019.

\bibitem{MaoZhaLet:16}
Y.~Mao, J.~Zhang, and K.~B. Letaief, ``Dynamic computation offloading for
mobile-edge computing with energy harvesting devices,'' \emph{{IEEE} J. Sel.
	Areas Commun.}, vol.~34, no.~12, pp. 3590--3605, Sep. 2016.

\bibitem{QQMaoZhaLet:17}
Y.~Mao, J.~Zhang, and K.~B. Letaief, ``Joint task offloading scheduling and
transmit power allocation for mobile-edge computing systems,'' in \emph{Proc.
	IEEE. Wireless Commun. Netw. Conf. (WCNC)}, San Francisco, CA, USA, Mar.
2017, pp. 1--6.

\bibitem{HuZonWanZhu:19}
H.~Hu, P.~Zong, H.~Wang, and H.~Zhu, ``Performance analysis for {D}2{D}-enabled
cellular networks with mobile edge computing,'' in \emph{Proc. Int. Conf.
	Wireless Commun. Signal Process. (WCSP)}, Xi'an, China, Oct. 2019, pp. 1--6.

\bibitem{ParLee:20}
C.~Park and J.~Lee, ``Mobile edge computing-enabled heterogeneous networks,''
\emph{{IEEE} Trans. Wireless Commun.}, vol.~20, no.~2, pp. 1038--1051, Oct.
2020.

\bibitem{KoHanHua:18}
S.-W. Ko, K.~Han, and K.~Huang, ``Wireless networks for mobile edge computing:
Spatial modeling and latency analysis,'' \emph{{IEEE} Trans. Wireless
	Commun.}, vol.~17, no.~8, pp. 5225--5240, Jun. 2018.

\bibitem{TraChaPom:19}
T.~X. Tran, K.~Chan, and D.~Pompili, ``Costa: Cost-aware service caching and
task offloading assignment in mobile-edge computing,'' in \emph{Proc. IEEE
	Int. Conf. Sens., Commun., Netw. (SECON)}, Boston, MA, USA, Jun. 2019, pp.
1--9.

\bibitem{LiZhaJiLi:19}
J.~Li, H.~Zhang, H.~Ji, and X.~Li, ``Joint computation offloading and service
caching for {MEC} in multi-access networks,'' in \emph{Proc. IEEE Int. Symp.
	Pers., Indoor Mob. Radio Commun. (PIMRC)}, Istanbul, Turkey, Sep. 2019, pp.
1--6.

\bibitem{CheHaoHuHosGho:18}
M.~Chen, Y.~Hao, L.~Hu, M.~S. Hossain, and A.~Ghoneim, ``Edge-{C}o{C}a{C}o:
Toward joint optimization of computation, caching, and communication on edge
cloud,'' \emph{IEEE Wireless Commun.}, vol.~25, no.~3, pp. 21--27, Jul. 2018.

\bibitem{ZhaHouWanCha:18}
T.~Zhao, I.-H. Hou, S.~Wang, and K.~Chan, ``Red/{L}e{D}: An asymptotically
optimal and scalable online algorithm for service caching at the edge,''
\emph{{IEEE} J. Sel. Areas Commun.}, vol.~36, no.~8, pp. 1857--1870, Jun.
2018.

\bibitem{WenCuiQueZheJin:20}
W.~Wen, Y.~Cui, T.~Q. Quek, F.-C. Zheng, and S.~Jin, ``Joint optimal software
caching, computation offloading and communications resource allocation for
mobile edge computing,'' \emph{{IEEE} Trans. Veh. Technol.}, vol.~69, no.~7,
pp. 7879--7894, May 2020.

\bibitem{CuiJia:16}
Y.~Cui and D.~Jiang, ``Analysis and optimization of caching and multicasting in
large-scale cache-enabled heterogeneous wireless networks,'' \emph{{IEEE}
	Trans. Wireless Commun.}, vol.~16, no.~1, pp. 250--264, Oct. 2016.

\bibitem{MarSyeKyuNikIliAntRahAse:18}
S.~Markov, N.~Kyurkchiev, A.~Iliev, and A.~Rahnev, ``On the approximation of
the generalized cut functions of degree p+ 1 by smooth hyper-log-logistic
function,'' \emph{Dyn. Syst. Appl.}, vol.~27, no.~4, pp. 715--728, Aug. 2018.

\bibitem{RazMeiHonMinLuoZhiPanJon:14}
M.~Razaviyayn, M.~Hong, Z.-Q. Luo, and J.-S. Pang, ``Parallel successive convex
approximation for nonsmooth nonconvex optimization,'' in \emph{Proc. Neural
	Inf. Process. (NIPS)}, Montreal, QC, Canada, Dec. 2014, pp. 1--9.

\bibitem{ElsSulAloWin:17}
H.~ElSawy, A.~Sultan-Salem, M.-S. Alouini, and M.~Z. Win, ``Modeling and
analysis of cellular networks using stochastic geometry: A tutorial,''
\emph{{IEEE} Commun. Surveys Tuts.}, vol.~19, no.~1, pp. 167--203, Nov. 2016.

\bibitem{MarBinHilHinKouMilUpt:02}
D.~T. Marr, F.~Binns, D.~L. Hill, G.~Hinton, D.~A. Koufaty, J.~A. Miller, and
M.~Upton, ``Hyper-threading technology architecture and microarchitecture.''
\emph{Intel Technol. J.}, vol.~6, no.~1, pp. 1--12, Feb. 2002.

\bibitem{XiaAoShaSha:20}
X.~Ma, A.~Zhou, S.~Zhang, and S.~Wang, ``Cooperative service caching and
workload scheduling in mobile edge computing,'' in \emph{Proc. IEEE Conf.
	Comput. Commun. (INFOCOM)}, Toronto, ON, Canada, Jul. 2020, pp. 2076--2085.

\bibitem{RomBen:20}
R.~Fantacci and B.~Picano, ``Performance analysis of a delay constrained data
offloading scheme in an integrated cloud-fog-edge computing system,''
\emph{IEEE Trans. Veh. Technol.}, vol.~69, no.~10, pp. 12\,004--12\,014, Oct.
2020.

\bibitem{QiaJeiXiaXia:20}
Q.~Kuang, J.~Gong, X.~Chen, and X.~Ma, ``Analysis on computation-intensive
status update in mobile edge computing,'' \emph{IEEE Trans. Veh. Technol.},
vol.~69, no.~4, pp. 4353--4366, Apr. 2020.

\bibitem{AbhSor:10}
A.~Abhari and M.~Soraya, ``Workload generation for youtube,'' \emph{Multimedia
	Tools and Applications}, vol.~46, no.~1, pp. 91--118, Jun. 2010.

\bibitem{FedOlaBeaPet:21}
F.~Chiariotti, O.~Vikhrova, B.~Soret, and P.~Popovski, ``Peak age of
information distribution for edge computing with wireless links,'' \emph{IEEE
	Trans. Commun.}, vol.~69, no.~5, pp. 3176--3191, May 2021.

\bibitem{ZinOsv:19}
J.~Zhang and O.~Simeone, ``On model coding for distributed inference and
transmission in mobile edge computing systems,'' \emph{IEEE Commun. Lett.},
vol.~23, no.~6, pp. 1065--1068, Jun. 2019.

\bibitem{QixJinLeiZhiZhuZhi:20}
Q.~Zhang, J.~Chen, L.~Ji, Z.~Feng, Z.~Han, and Z.~Chen, ``Response delay
optimization in mobile edge computing enabled uav swarm,'' \emph{IEEE Trans.
	Veh. Technol.}, vol.~69, no.~3, pp. 3280--3295, Mar. 2020.

\bibitem{XinZheYuaXin:21}
X.~Qin, Z.~Song, Y.~Hao, and X.~Sun, ``Joint resource allocation and trajectory
optimization for multi-uav-assisted multi-access mobile edge computing,''
\emph{IEEE Commun. Lett.}, vol.~10, no.~7, pp. 1400--1404, Jul. 2021.

\bibitem{ChaKaiHyuByo:16}
C.~You, K.~Huang, H.~Chae, and B.-H. Kim, ``Energy-efficient resource
allocation for mobile-edge computation offloading,'' \emph{IEEE Transactions
	on Wireless Communications}, vol.~16, no.~3, pp. 1397--1411, Mar. 2016.

\bibitem{NovDhiAnd:13}
T.~D. Novlan, H.~S. Dhillon, and J.~G. Andrews, ``Analytical modeling of uplink
cellular networks,'' \emph{{IEEE} Trans. Wireless Commun.}, vol.~12, no.~6,
pp. 2669--2679, May 2013.

\bibitem{SinDhiAnd:13}
S.~Singh, H.~S. Dhillon, and J.~G. Andrews, ``Offloading in heterogeneous
networks: Modeling, analysis, and design insights,'' \emph{{IEEE} Trans.
	Wireless Commun.}, vol.~12, no.~5, pp. 2484--2497, Apr. 2013.

\bibitem{Kle:75}
L.~Kleinrock, ``Theory, volume 1, queueing systems,'' 1975.

\bibitem{FraJan:01}
G.~J. Franx, ``A simple solution for the m/d/c waiting time distribution,''
\emph{Oper. Res. Lett.}, vol.~29, no.~5, pp. 221--229, Dec. 2001.

\bibitem{BoyVan:B04}
S.~Boyd and L.~Vandenberghe, \emph{Convex Optimization}.\hskip 1em plus 0.5em
minus 0.4em\relax Cambridge, UK: Cambridge University Press, 2004.

\bibitem{BlaGio:15}
B.~Blaszczyszyn and A.~Giovanidis, ``Optimal geographic caching in cellular
networks,'' in \emph{Proc. IEEE Int. Conf. Commun. (ICC)}, London, UK, Jun.
2015, pp. 1--6.

\bibitem{ThiAnPha:18}
H.~A. Le~Thi and T.~P. Dinh, ``{DC} programming and {DCA}: Thirty years of
developments,'' \emph{Math. Prog.}, vol. 169, no.~1, pp. 5--68, Jan. 2018.

\bibitem{SriLan:09}
B.~K. Sriperumbudur and G.~R. Lanckriet, ``On the convergence of the
concave-convex procedure,'' in \emph{Proc. Neural Inf. Process. Syst.
	(NIPS)}, Vancouver, BC, Canada, Dec. 2009, pp. 1--9.

\bibitem{BiHuaZha:20}
S.~Bi, L.~Huang, and Y.-J.~A. Zhang, ``Joint optimization of service caching
placement and computation offloading in mobile edge computing systems,''
\emph{{IEEE} Trans. Wireless Commun.}, vol.~19, no.~7, pp. 4947--4963, Apr.
2020.

\bibitem{RenYuHeLi:19}
J.~Ren, G.~Yu, Y.~He, and G.~Y. Li, ``Collaborative cloud and edge computing
for latency minimization,'' \emph{{IEEE} Trans. Veh. Technol.}, vol.~68,
no.~5, pp. 5031--5044, Mar. 2019.

\bibitem{kwak2018hybrid}
J.~Kwak, Y.~Kim, L.~B. Le, and S.~Chong, ``Hybrid content caching in 5{G}
wireless networks: Cloud versus edge caching,'' \emph{{IEEE} Trans. Wireless
	Commun.}, vol.~17, no.~5, pp. 3030--3045, May 2018.

\bibitem{TamBenNarLat:15}
S.~Tamoor-ul Hassan, M.~Bennis, P.~H. Nardelli, and M.~Latva-Aho, ``Modeling
and analysis of content caching in wireless small cell networks,'' in
\emph{Proc. IEEE Int. Symp. Wireless Commun. Syst. (ISWCS)}, Brussels,
Belgium, Aug. 2015, pp. 1--5.

\bibitem{BasBenDeb:15}
E.~Ba{\c{s}}tu{\u{g}}, M.~Bennis, and M.~Debbah, ``A transfer learning approach
for cache-enabled wireless networks,'' in \emph{Proc. Int. Symp. Modeling
	Optim. Mob., Ad Hoc, Wireless Netw. (WiOpt)}, Mumbai, India, May 2015, pp.
1--6.
\end{thebibliography}
\end{document}